\documentclass[12pt,showpacs,prc]{revtex4}
\usepackage{epsfig,amssymb,amsfonts}
\usepackage{graphicx}
\begin{document}

\draft

\title{Multiplicity Distributions of Intermediate Mass Fragments in the Thermodynamic Model}
 
\author{L. Shi, S. Das Gupta}

\address{Physics Department, McGill University, 
Montr{\'e}al, Canada H3A 2T8}

\date{\today}

\begin{abstract}
{
Multiplicity distributions of intermediate mass fragments (IMF) seen in intermediate energy heavy ion collisions have been of great interest in the last ten years.  The distributions of single intermediate sized element are close to Poissonian and have been subjected to widely different physical interpretations.  We calculate the multiplicity distributions in a 
thermodynamic model whose physical basis is equal population in 
phase space which implies equilibrium statistics. We found distributions of single element are to a good approximation Poissonian, and such result is attributed to the small emission probability of a given IMF among all the possible channels. We also found significant deviation from Poisson distribution for a range of IMFs, and suggest strong correlations as an explanation. Systematics of the multiplicity distributions are explored in both one component and two component models. In one component model, sudden changes for the multiplicity distributions are found at the phase transition region. In two component model, the multiplicity distributions have more subtleties.
Results from the current model are compared with available data, and good agreements are found. New signals for the multi-particle correlations are also suggested. 
}
\end{abstract}

\maketitle 

\section{Introduction}
The multiplicity distributions of intermediate mass fragments ($Z$=3 to 20) seen in intermediate energy heavy ion collisions seem to be particularly simple.  They closely resemble 
Binomial distribution.  The Binomial distribution has two 
parameters $m$ and $p$ and in the limit $m \rightarrow \infty$, $p\rightarrow 0$ and $mp \rightarrow \langle N \rangle$ (where $\langle N \rangle$ is a constant) goes over to a Poissonian.  Experimental data resemble a Poissonian \cite{Beaulieu:1998a,Beaulieu:1998b,Moretto:1995}. This distribution has been the subject of widely different interpretations \cite{Beaulieu:1998a,Beaulieu:1998b,Moretto:1995,Moretto:1997,Tsang:1998,Toke:1997,Skulski:1998,Bauer:1999}.

In theoretical models the multiplicity distributions can be very difficult to calculate.  BUU (Boltzmann-Uehling-Uhlenbeck) simulations can not give accurate predictions for fragment 
productions \cite{Bertsch:1988}.  In the Copenhagen model \cite{Bondorf:1995,Bondorf:1985a,Bondorf:1985b,Bondorf:1985c} or Berlin 
model \cite{Gross:1997,Gross:1985,Gross:1987a,Gross:1987b} or in the lattice gas model \cite{Yang:1952,Pan:1995a,Pan:1995b,Campi:1997,Chomaz:1998tp,Gulminelli:1999,Carmona:1998,Dasgupta:2000} events are generated by Monte-Carlo simulations and generating a multiplicity distribution would be extremely time consuming. On the other hand, new techniques have been formulated where, assuming thermodynamic equilibrium, various quantities of physical interest can be calculated bypassing lengthy Monte-Carlo simulations \cite{Chase:1993nw,Chase:1995ad,Das:2003}.  Except for some simplifications, the physical basis of the model is the same as the Copenhagen or Berlin model.
We therefore embarked upon calculating the multiplicity distributions in the thermodynamic model.

The plan of the paper is as follows. A brief description of the canonical model is given in Sect.\ref{section:model}. Sect.\ref{section:multiplicity} is devoted to the multiplicity distributions of single or a range of intermediate sized clusters in one and two component model. Since decay and feeding process could also affect the multiplicity distributions detected in experiments, we also discussed them in the same section. Multiplicity distributions are strongly correlated with the system size, and the source size dependence is discussed in Sect.\ref{sect:systemsize}. The system size dependence of multiplicity distributions is compared with data in the same section.
The multiplicity gated charge yield is related to the correlation (and anti-correlation) between IMFs and other clusters, and is discussed in Sect.\ref{sect:gatedyield}. The variance of multiplicity distributions is one of the most simple observables for cluster correlations, and is discussed in Sect.\ref{sect:varianceandmean}. Comparison with data on variance is also made in the same section. In the end, we summarize the results in Sect.\ref{sect:summary}. The two appendices show the multiplicity tagging method for the multiplicity distributions, and a unified method for decay and feeding in the current model.

\section{The Thermodynamic Model}
\label{section:model}

The thermodynamic model has been described in many places 
\cite{Chase:1993nw,Chase:1995ad,Das:2003,Dasgupta:2001,Bhattacharyya:1999,Dasgupta:1998}.  For completeness, to enumerate the parameters and the techniques needed to calculate the multiplicity distributions, we provide some details.  We describe first the model for one kind of particles.  Although unrealistic, the model brings out, with least complications, the physics of the situation readily.  The generalization to two kinds is done next 
and follows the procedures given in \cite{Bhattacharyya:1999,Dasgupta:2001}.

If there are $A$ identical particles of only one kind in an enclosure at temperature $T$, the partition function of the system can be written as 
\begin{eqnarray} 
Q_A=\frac{1}{A!}(\omega)^A \, ,
 \label{eq:sect2:eq1}
\end{eqnarray} Here $\omega$ is the partition function of one particle.  For a spinless particle this is $\omega=\frac{V}{h^3}(2\pi mT)^{3/2}$; $m$ is the mass 
of the particle; $V$ is the available volume within which each particle moves; $A!$ corrects for Gibb's paradox.  If there are many species, the generalization is \begin{eqnarray} 
Q_A=\sum\prod_i\frac{(\omega_i)^{n_i}}{n_i!}\, \delta(\sum_i i \times n_i - A) \, ,
 \label{eq:sect2:eq2}
\end{eqnarray}
Here $\omega_i$ is the partition function of a composite which has $i$ nucleons.  For a dimer $i=2$, for a trimer $i=3$ etc.  Eq. (\ref{eq:sect2:eq2}) is no longer trivial to calculate.  The trouble is with the sum in the right hand side of Eq. (\ref{eq:sect2:eq2}).  The sum is restrictive.  We need to consider only those partitions of the number $A$ which satisfy $A=\sum i\ n_i$.  The number of partitions which satisfies the sum is enormous.  We can call a given allowed partition to be a channel.  The probability of the occurrence of a given channel $P(\vec n)\equiv P(n_1,n_2,n_3....)$ is 
\begin{eqnarray} 
P(\vec n)=\frac{1}{Q_A}\prod\frac{(\omega_i)^{n_i}}{n_i!} \,.
 \label{eq:sect2:eq3}
\end{eqnarray}
The average number of composites of $i$ nucleons is easily seen from the above equation to be 
\begin{eqnarray}
 \langle n_i \rangle =\omega_i\frac{Q_{A-i}}{Q_A} \, .
 \label{eq:sect2:eq4}
\end{eqnarray}
Since $\sum in_i=A$, one readily arrives at a recursion relation 
\cite{Dasgupta:1998}
\begin{eqnarray} 
Q_A=\frac{1}{A}\sum_{k=1}^{k=A}k\omega_kQ_{A-k} \, .
 \label{eq:sect2:eq5}
\end{eqnarray}
For one kind of particle, $Q_A$ above is easily evaluated on a computer for 
$A$ as large as 3000 in matter of seconds.  It is this recursion relation that makes the computation so easy in the model.  Of course, once one has the partition function all relevant thermodynamic quantities can be 
computed.

We now need an expression for $\omega_k$ which can mimic the nuclear physics situation.  We take 
\begin{eqnarray} 
\omega_k=\frac{V}{h^3}(2\pi mT)^{3/2}\times q_k  \, ,
  \label{eq:sect2:eq6}
\end{eqnarray} 
where the first part arises from the centre of mass motion of the composite which has $k$ nucleons and $q_k$ is the internal partition function. For $k=1$, $q_k=1$ and for $k\ge 2$ it is taken to be 
\begin{eqnarray} 
q_k=\exp[(W_0k-\sigma(T)k^{2/3}+T^2k/\epsilon_0)/T] \,,
 \label{eq:sect2:eq7}
\end{eqnarray}
Here, as in \cite{Bondorf:1995}, $W_0$=16 MeV is the volume energy term, 
$\sigma(T)$
is a temperature dependent surface tension term and the last term arises from summing over excited states in the Fermi-gas model.  The value of $\epsilon_0$ is taken to be 16 MeV.  The explicit expression for $\sigma(T)$ used here is $\sigma(T)=\sigma_0[(T_c^2-T^2)/(T_c^2+T^2)]^{5/4}$ with $\sigma_0=$18 MeV and $T_c=18$ MeV. In the nuclear case one might be tempted to interpret $V$ of Eq.(\ref{eq:sect2:eq6}) as simply the freeze-out volume but it is clearly less than that; $V$ is the volume available to the particles for the center of mass motion.  Assume that the only interaction between clusters is that they can not overlap one another then in the Van der Waals spirit we take $V=V_{freeze}-V_{ex}$ where $V_{ex}$ is taken here to be constant and equal to $V_0=A/\rho_0$. 

Our objective is to calculate the probability that in an event, the probability that $n$ clusters of a composite which has $k$ nucleons 
is emitted.  This is given by
\begin{eqnarray} 
P_n(k)=\frac{(\omega_k)^n}{n!}\frac{Q_{A-nk}(\omega_k=0)}{Q_A}  \,,
 \label{eq:sect2:eq8}
\end{eqnarray}
Here $Q_{A-nk}(\omega_k=0)$ is the partition function of $A-nk$ particles but with the restriction that there are no composites of $k$ nucleons. That is 
\begin{eqnarray} 
Q_{A-nk}(\omega_k=0)=\sum\prod_{i\ne k}\frac{(\omega_i)^{n_i}}{n_i!} \,.
 \label{eq:sect2:eq9} 
\end{eqnarray} 
This can also be obtained by calculating the partition function for $A-nk$ particles as before where all $\omega_i$'s are the same except for $\omega_k$ which is set to zero.

We will also be interested in a more complicated situation.  We will want to have the probability that $n$ intermediate mass fragments are emitted where the intermediate mass fragments span a range of composites.  Here we take this range to be 6 to 20.  Thus $n=\sum_{\alpha} n_i$ where $\alpha$ is the group 6 to 20.  The algebra needed to calculate this is given in Appendix \ref{appendix:tag}.

The decay of these clusters formed during the multifragmentation process will inevitably change the multiplicity distributions. Since the experiment will only observe the multiplicity distribution after the decay process, it is necessary to include the effects of cluster decay process in our considerations. The algebra is readily worked out in Appendix \ref{appendix:decay}. The essential idea is to group the after-decay clusters into two groups, one with all particles that still stay in the group we are interested in and another with all particles that go out of the group we are interested in, each clusters in the two group have a modified formation probability. Then since we are only interested in one of the group, we will only tag the group of interest, and leave the other group untagged. In the end, we only need to redo the partition with a modified formation probability and we can find the exact solution for the decay process.
The same idea also applies to the feeding from other clusters into the group of interest, with only a change of decay probability $\epsilon \leftrightarrow (1-\epsilon)$, and a redefinition of the group of interest. Details of the algebra are in appendix \ref{appendix:decay}.

These considerations can be extended to two kinds of particles, neutrons and protons \cite{Bhattacharyya:1999,Dasgupta:2001,Souza:2003}.  Now a composite is labelled by two indices $i \rightarrow (A_i, Z_i)$, and total system is also labelled by two indices $A \rightarrow (A, Z)$.  The partition function for a system with $Z$ protons and $A-Z$ neutrons is given by 
\begin{eqnarray} 
Q_{A,Z}=\sum\prod_{i}\frac{(\omega_{i})^{n_{i}}}{n_{i}!} \, \delta(\sum_i A_i \times n_i-A)\delta(\sum_i Z_i \times n_i - Z) \, ,
 \label{eq:sect2:eq10} 
\end{eqnarray}
There are two constraints: $A=A_i\times n_{i}$ and $Z=\sum Z_i\times n_{i}$ . These lead to two recursion relations any one of which can be used. For example \begin{eqnarray} 
Q_{A,Z}=\frac{1}{Z}\sum_{i}Z_i\omega_{i}Q_{A-A_i,Z-Z_i} \,,
 \label{eq:sect2:eq11}
\end{eqnarray}
where
\begin{eqnarray}
\omega_{i}=\frac{V}{h^3}(2\pi mT)^{3/2}(A_i)^{3/2}\times q_{i}  \,,
 \label{eq:sect2:eq12}
\end{eqnarray} 
Here $q_{i}$ is the internal partition function.  These could be taken from experimental binding energies, excited states and some model for the continuum or from the liquid drop model in combination with other models. The versatility of the model lies in being able to accommodate any choices for $q_{i}$.  A choice of $q_{i}$ from a combination of the liquid drop model for binding energies and the Fermi-gas model for excited states that has been used is 
\begin{eqnarray} 
q_{i}=\exp\left\{\frac{1}{T}\left[W_0 A_i-\sigma (A_i)^{2/3}-\kappa\frac{(Z_i)^2}
{(A_i)^{1/3}}-s\frac{(A_i-2Z_i)^2}{A_i}+\frac{T^2 A_i}{\epsilon_0}\right]\right\} \, .
 \label{eq:sect2:eq13} 
\end{eqnarray}
One readily recognizes in the parameterization above the volume term, the surface term, the Coulomb energy term, the symmetry energy term and contribution from the excited state. Coulomb repulsion between composites is sometimes taken into account by Wigner-Seitz approximation \cite{Bondorf:1995}.

\section{multiplicity distribution}
\label{section:multiplicity}

\subsection{one component system}
\label{section:multiplicity:subsect:onecomp}

The multiplicity distribution of intermediate sized clusters (6-40) is our primary interest, because these clusters are abundantly produced and are believed to be directly produced from multifragmentation process \cite{Bondorf:1985a,Bondorf:1985b,Bondorf:1985c,Bondorf:1995,Gross:1985,Gross:1987a,Gross:1987b,Gross:1997}. Lighter clusters have multiple origins. Heavier clusters are often less abundant and more difficult to identify in experiments \cite{D'Agostino:1999is,Liu:2002tp}. 
The essential features for multiplicity distribution is already present in a single component model, and we will start from this model first. We are interested in the multiplicity distribution of single sized clusters as well as the multiplicity distribution of a range of clusters.     

\begin{figure}[tbph!]
\center
\includegraphics[scale=0.4]{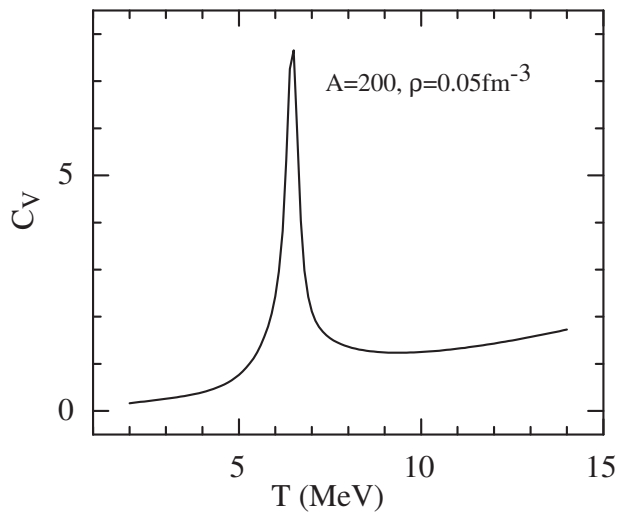}  % {m01-00.eps}
\caption{Specific heat $C_V$ as a function of temperature for a system of total mass $A=200$. The density of the fragmentation source is fixed at $\rho=0.05fm^{-3}$. The peak at $6.5$ MeV is defined as the phase transition temperature $T_b$ in the current one component canonical model.} 
 \label{fig:comp1:cv-peak}
\end{figure}

As is well known, the current canonical model exhibits features of a first order phase transition \cite{Das:2003,Bugaev:2000ar}. The divergence in the specific heat at the phase transition temperature is replaced by a finite peak due to finite system size effect, and the phase transition temperature $T_b$ is shifted to a lower value \cite{Das:2003}. The specific heat per particle is plotted as a function of system size in Fig.\ref{fig:comp1:cv-peak}. The temperature at which the specific heat is maximum is called phase transition temperature $T_b$ for the finite system. As will be shown later, some of the fragmentation characteristics change as the temperature of the fragmentation source changes from below phase transition temperature $T<T_b$ to above phase transition temperature $T>T_b$. Thus, the phase transition temperature provides a natural temperature scale for the fragmentation process in the current model.

\begin{figure}[tbph!]
\center
\includegraphics[scale=0.4]{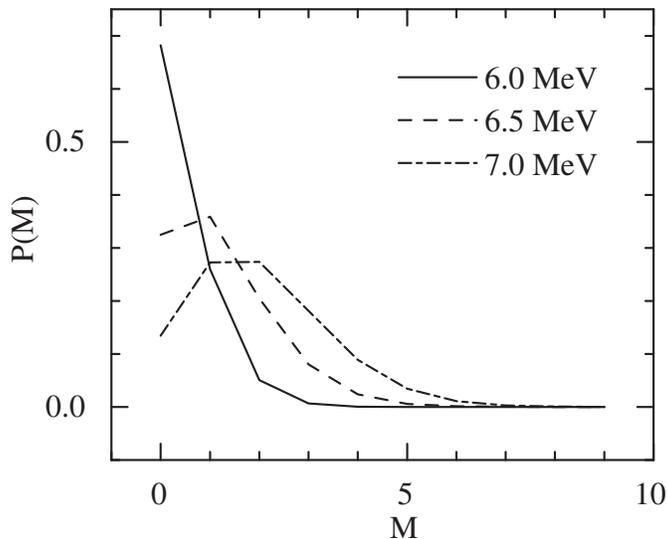}     %{m02-00b.eps}
\caption{The multiplicity distribution $P(M)$ of a single cluster with mass $k=6$ as a function of multiplicity of this cluster $M$ in one component canonical model. The fragmentation source has 200 particles, and density at fragmentation is $0.05fm^{-3}$. The phase transition temperature for this source is at $T_b=6.5$ MeV. The left, middle and right panels are for temperatures of $T=6$, $6.5$, and $7$ MeV respectively.} 
 \label{fig:comp1:mdist-singles}
\end{figure}

Fig.\,\ref{fig:comp1:mdist-singles} shows the multiplicity distribution of a single cluster with mass $k=6$, at three different temperatures $T<T_b$, $T \approx T_b$, and $T>T_b$. One of the most striking features in Fig.\,\ref{fig:comp1:mdist-singles} is that the probability of high multiplicity event is exponentially small. The Poisson distribution with same mean as the corresponding canonical distribution can be compared with the canonical results. In the linear scale used in Fig.\,\ref{fig:comp1:mdist-singles}, it is quite difficult to see any difference between the Poisson distribution and the canonical results for multiplicity distribution. We have tried to fit the canonical result with either a Poisson distribution or a binomial distribution as suggested by Merotto {\em et al.} \cite{Moretto:1995}, the fit is always very good. The reason for a good fit is easy to understand. The canonical results for multiplicity distribution have practically only a few relevant terms, probability for all high multiplicity events are exponentially small. So the fitting essentially only need to fit the first few terms of the multiplicity distribution, and the higher multiplicity terms are just too small to affect the fit. In such fitting, the condition probability distribution sums up to unity is automatically satisfied by the fit distributions. 

In the case of $T=6.0$ MeV in Fig.\,\ref{fig:comp1:mdist-singles}, we can safely neglect terms with $M \geq 3$, and we are left with only two significant terms in the distribution. Then it is not a surprise to see a good fit with Poisson distribution. Since Poisson distribution has only one parameter, could fit well the canonical result, we will hereafter only compare the results of the canonical model with the Poisson distribution. Even at $T=7$ MeV, the multiplicity distribution of a single cluster $k=6$ is well fitted by a simple Poisson distribution. And this result holds for effectively all small clusters as long as the system size is considerably larger than the cluster we are interested in. This seemingly simple distribution has been the subject of various interpretations \cite{Beaulieu:1998a,Beaulieu:1998b,Moretto:1995,Moretto:1997,Tsang:1998,Toke:1997,Skulski:1998,Bauer:1999}.

In fact, the good fit with a simple Poisson distribution is not a coincidence. In a grand canonical model for multifragmentation process, the single cluster multiplicity distribution is always a Poisson distribution. The total particle number is not conserved in the grand canonical case, and a production of a certain cluster does not affect the yield of a second similar cluster at all \cite{Bugaev:2000ar,Das:2001}. And each of these processes are simply combined with a factorial factor accounting for statistics, which then yield the Poisson distribution. For a canonical result to yield a Poisson distribution, we need to require small probability for single cluster formation, so that formation of the first cluster will not significantly affect the formation of a second same cluster. This condition is fulfilled when the cluster is much smaller than the entire fragmentation source, and there are many similar channels for particles to go into beside the one we are interested. In other words, the clusters in the low multiplicity events are effectively produced independently and little correlation exists between them. Such independent production mechanism together with the factorial factor for statistics always yields a near Poisson distribution.

\begin{figure}[tbph!]
\center
\includegraphics[scale=0.4]{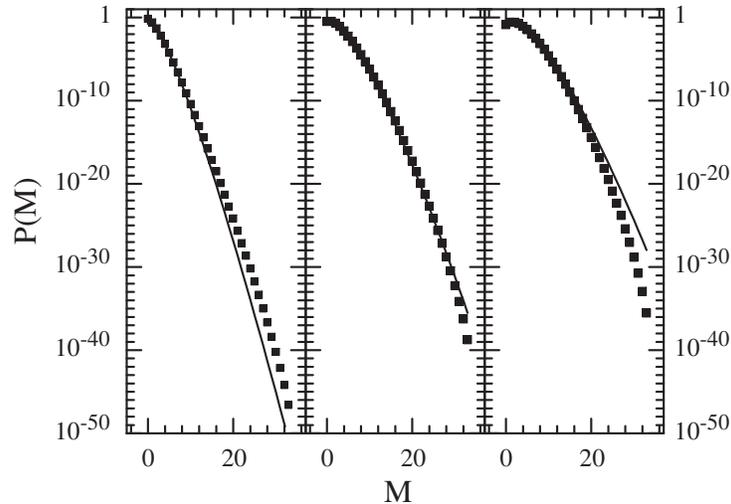}   %{m03-00.eps}
\caption{Same as \ref{fig:comp1:mdist-singles}. The multiplicity distribution $P(M)$ of single clusters, $k=6$, as a function of multiplicity $M$ in one component canonical model, in a log plot. The left, middle and right panels are for temperatures of $T=6$, $6.5$, and $7$ MeV respectively. In all three panels, the canonical results are shown with square symbols and the Poisson distribution with the same mean is shown with solid line. It is clear that canonical result for a single cluster follows closely Poisson distribution, except at large multiplicities. However, significant difference shows up at large multiplicities. There is a change of 
super-Poissonian distribution to sub-Poissonian distribution when temperature changes from below phase transition temperature to above phase transition temperature.} 
 \label{fig:comp1:mdist-singles-log-pstest}
\end{figure} 

Despite the good fit with a Poisson distribution, there are still significant differences in the canonical results and the Poisson distributions. If we look at the fine details of the multiplicity distribution and the Poisson distribution, we find the tails of the distributions differ considerably. In the log plot of the multiplicity distributions, Fig.\,\ref{fig:comp1:mdist-singles-log-pstest}, the canonical results show systematic deviation from the Poisson distribution. At below phase transition temperatures, the tail of the canonical result is significantly larger than the corresponding Poisson distribution with the same mean. Such a trend reverses at temperature above the phase transition temperature. Right at phase transition temperature, $T=T_b$, the distribution follows very closely the corresponding Poisson distribution.

At below $T_b$, the system is in liquid-gas coexistence phase. The particles in the system could either choose to reside in the the small clusters (gas phase) or in the large clusters (liquid phase), and thus incur large fluctuations in the distribution. Above phase transition temperature, $T>T_b$, the system is in gas phase. The particles are partitioned only in small clusters. The width of the multiplicity distribution should be reduced due to finite system size constraint.
At phase transition temperature, $T=T_b$, the different factors compete, and we are left with an almost Poisson distribution. So at the phase transition temperature, different constraints on the multiplicity distribution cancel, the system reaches a critical point. The small clusters suddenly do not feel the boundary of the system, and form according to the unconstrained distribution, that is, the Poisson distribution.

\begin{figure}[tbph!]
\center
\includegraphics[scale=0.6]{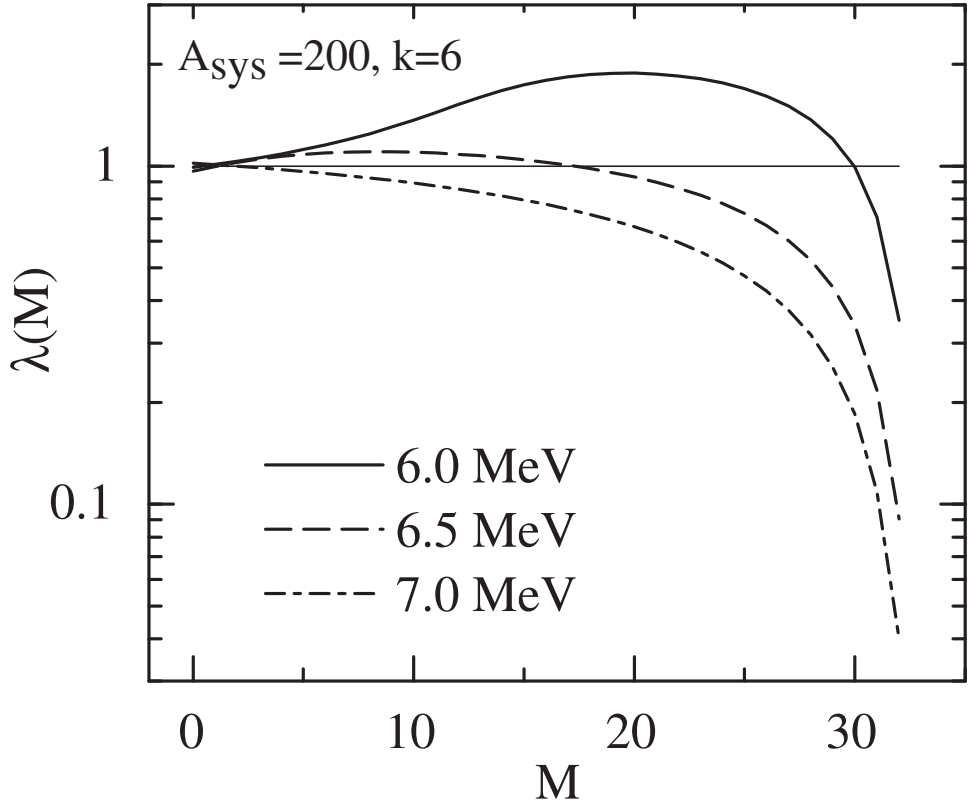}       % {m04-01.eps}
\caption{The Poisson test parameter $\lambda(M)$ for multiplicity distribution of single clusters, $k=6$, as a function of multiplicity $M$ in one component canonical model. See Eq.(\ref{eq:sect3:eq1}) for the definition of $\lambda(M)$.
By definition, a Poisson distribution will have $\lambda(M) \equiv 1$ for all $M$. The
Poisson test parameter shows the difference in the multiplicity distribution as the temperature changes from below to above $T_b$. Right at around phase transition temperature, the Poisson test parameter is most close to unity for a large range of multiplicities. While below and above phase transition temperature, the Poisson test parameter is far than from unity for most multiplicities.} 
 \label{fig:comp1:mdist-singles-pstest}
\end{figure}

To get more quantitative description of the differences between the canonical result and the Poisson distribution, it is instructive to define a Poisson test parameter
\begin{eqnarray}
\lambda(M) = \frac{(M+1) \times P(M+1) } {\langle M \rangle \times P(M) } \,.       
  \label{eq:sect3:eq1}
\end{eqnarray}
Where $P(M)$ is the probability of the multiplicity $M$ event, and $\langle M \rangle$ is the mean multiplicity.   
Such a ratio was introduced in \cite{Beaulieu:1998a} in the discussion of multiplicity distributions. For Poisson distribution, this test parameter is always exactly one. For canonical results, this test parameter scales out the obvious power decay factor, and shows the fine details of deviation from Poisson distribution without reference to any artificially constructed Poisson distributions.

The Poisson test parameter $\lambda(M)$ for the canonical results are plotted in Fig.\,\ref{fig:comp1:mdist-singles-pstest}. With multiplicity distributions such as obtained for $k=6$, the deviation from exact Poisson distribution is mostly in the high multiplicity tail. The Poisson test parameter reveals the quantitative deviation from Poisson distribution and such difference is readily seen at the high multiplicity tail. When we are dealing with a single sized cluster formed in the fragmentation process, $P(M)$ reflects not only the single formation factor associated with the production of such cluster, it also is constrained by the weight factor for phase space of other clusters. With a change of $M \rightarrow M+1$, as we can see from Eq.(\ref{eq:sect2:eq8}), the first weight factor related to the formation of the interested cluster changes $\omega^M / M! \rightarrow \omega^{(M+1)} / (M+1)!$, and is exactly cancelled by the ratio Eq.(\ref{eq:sect3:eq1}); the second weight factor related to the partition of the rest of the particles in the system also changes, $Q_{A-k \times M}(\omega_k=0) \rightarrow Q_{A-k \times (M+1)}(\omega_k=0)$. When the Poisson test parameter $\lambda(M)$ is nearly constant for small $M$, we infer that there is a rough scaling in the second partition weight when multiplicity $M$ changes.

The second weight factor is just the partition of $(A-k \times M)$ particles with the condition of not producing the pre-selected cluster $k$. With the understanding that the formation of the pre-selected cluster is one of the many channels the system could select, ignoring such a channel will not significantly change the total phase space. Then it became clear that the scaling in $Q_{A-k \times M}$ simple suggests a system size scaling in the total partition function. 
At any given temperature, the partition pattern is not expected to change significantly when system size changes by a small number as long as $k \times M << A$, so that the partition function will scale roughly with the system size. Such a scaling is of course going to break down when the size $(A-k \times M)$ does not remain large anymore, and thus finite system size effect kicks in.

\begin{figure}[tbph!]
\center
\includegraphics[scale=0.7]{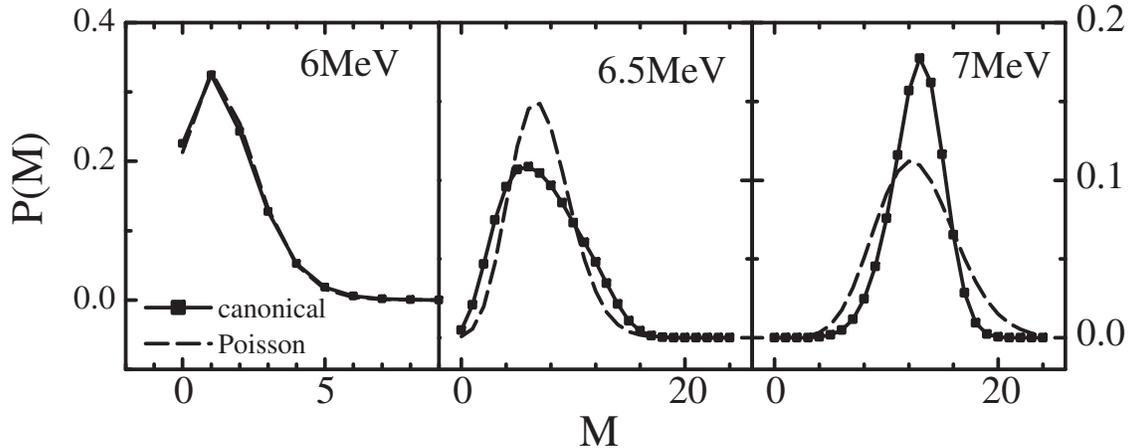}     %{m05-02.eps}
\caption{The multiplicity distribution $P(M)$ of intermediate mass fragments, $k=6-40$, as a function of multiplicity $M$ in the one component canonical model. The fragmentation source has total mass $200$, and its density is fixed at $0.05fm^{-3}$.
The left, middle and right panels correspond to temperatures below, at and above phase transition temperature. The Poisson distribution with the same mean is plotted together with the corresponding canonical distribution. Clearly, the deviation from Poisson distribution is visible even in this linear plot.
The mean multiplicity for a range of clusters is considerably higher than that for a single cluster.} 
 \label{fig:comp1:mdist-range-linear}
\end{figure}

\begin{figure}[tbph!]
\center
\includegraphics[scale=0.4]{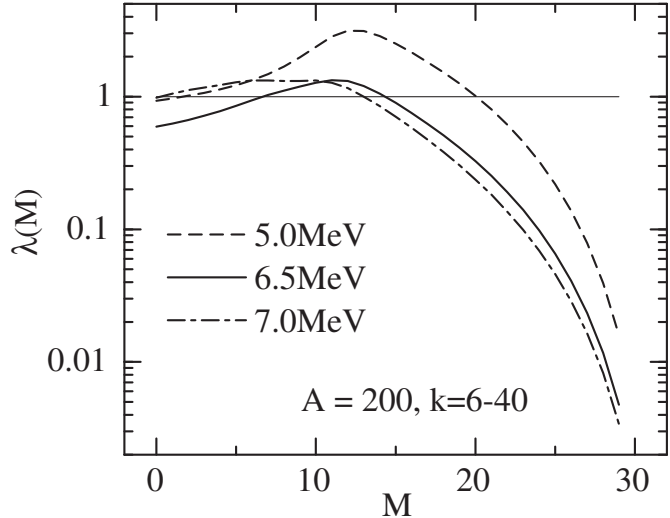}       % {m06-00.eps}
\caption{The Poisson test parameter for the multiplicity distribution of a range of clusters, k=6-40, in the one component canonical model. The system size and breakup volume are the same as in Fig.\ref{fig:comp1:mdist-range-linear}.} 
 \label{fig:comp1:mdist-range-pstest}
\end{figure}

While the production of a single cluster $k$ is only a small fraction of the phase space, the production of a range of clusters is a much more sizable fraction of the phase space, and could reveal the bulk structure of the partition phase space. For single component model, we take the intermediate mass fragment (IMF) to be $6 \leq k \leq 40$, and calculated the IMF multiplicity distribution, see Figs.\ref{fig:comp1:mdist-range-linear} and \ref{fig:comp1:mdist-range-pstest}. 

At below phase transition temperature, $T<T_b$, the multiplicity distribution for a range of IMFs is similar to the single cluster case because the the production of any IMF is small, and we find the Poisson fit is quite good. But when the IMFs are abundantly produced, $T \geq T_b$, the Poisson distribution does not describe the canonical result at all. The narrowing of the IMF multiplicity distribution seen at temperatures $T > T_b$ is significant compared with the multiplicity distribution of a single sized cluster. Such narrow distribution reveals the strong correlations between the IMF clusters. Also the quick change from a fairly good Poisson distribution to a strong sub-Poisson distribution is a manifestation of the strong correlations between IMF clusters.   

The Poisson test parameter for the IMF multiplicity distribution is plotted in Fig.\ref{fig:comp1:mdist-range-pstest}. Even though the distribution is not even close to Poisson distribution at temperatures $T \geq T_b$, the Poisson test parameter still reveals interesting features of the distribution. In the low multiplicity region, $M <10$, the P(M)'s at $T=T_b$ is significantly lower at either $T<T_b$ or $T>T_b$. Such a trend is not see in the multiplicity distribution for single IMF cluster in Fig.\ref{fig:comp1:mdist-singles-pstest}. If we look more carefully at the Poisson test parameter for M=0 in Fig.\ref{fig:comp1:mdist-singles-pstest}, it is clear that the reduction of $\lambda(M=0)$ at $T=6.5$ MeV is also present, but not as prominent as in Fig.\ref{fig:comp1:mdist-range-pstest}. Here, we find that a small signal in multiplicity distributions is enhanced by the selection of a range of IMFs, and we will also come to similar conclusions later in Sects.\ref{sect:gatedyield} and \ref{sect:varianceandmean}. For high multiplicities, $M>15$, the lines for all three temperatures are very similar.  The most likely clusters will be small clusters if the multiplicity is high, thus the large multiplicity tail of the distribution will gradually resembles the multiplicity distribution for $A=6$. The three curves are almost parallel at very high multiplicities, which also suggests the common origin for the high multiplicity events.

\subsection{two component system}

For two component canonical model, we can also use the tagging method to find the multiplicity distribution for a single isotope or for a range of isotopes. The production of a single light isotope is again a small fraction of the phase space, multiplicity higher than $1$ event will be only a small faction of the total fragmentation phase space. But if such rare event do occur, it may present different systematics due to the additional isospin degree of freedom. In the two component model, the partition function has two constraints, one for system size and another for charge, or isospin asymmetry. On the other hand, the available clusters has one more label for isospin too. For each fragment of given size, there are a range of clusters with different isospin asymmetry.

\begin{figure}[tbph!]
\center
\includegraphics[scale=0.4]{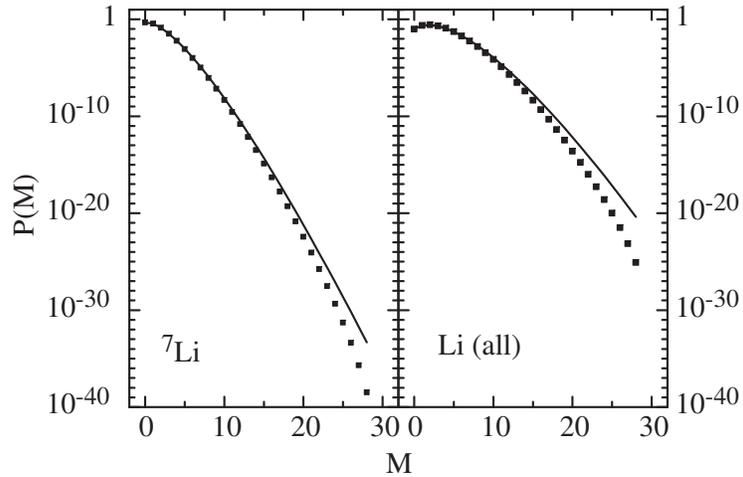}       % {mm-00.eps}
\caption{The multiplicity distribution of $Li7$ and $Li$(all) at $T=5.2$MeV, in two component model with $A_{sys}=200$ and $Z_{sys}=86$, in log scale. The data symbols are the results from the canonical model while solid lines indicate the corresponding Poisson distribution with same mean.}  
 \label{fig:comp2:mdist-single-Li}
\end{figure}

%\begin{figure}[tbph!]
%\center
%\includegraphics[scale=0.6]{m7-00.eps}
%\caption{The Poisson test parameter for multiplicity distribution of single clusters, $Z=3$, $A=5,6,7,8,9$, in two component model. The system is $(A_{sys}=200,Z_{sys}=86)$ $T=5.2$ MeV while phase transition temperature is $T_b=4.6 MeV$ for this system. The line at $\lambda(M)=1$ is for exact Poisson distribution.} 
% \label{fig:comp2:mdist-single-pstest}
%\end{figure}

\begin{figure}[tbph!]
\center
\includegraphics[scale=0.6]{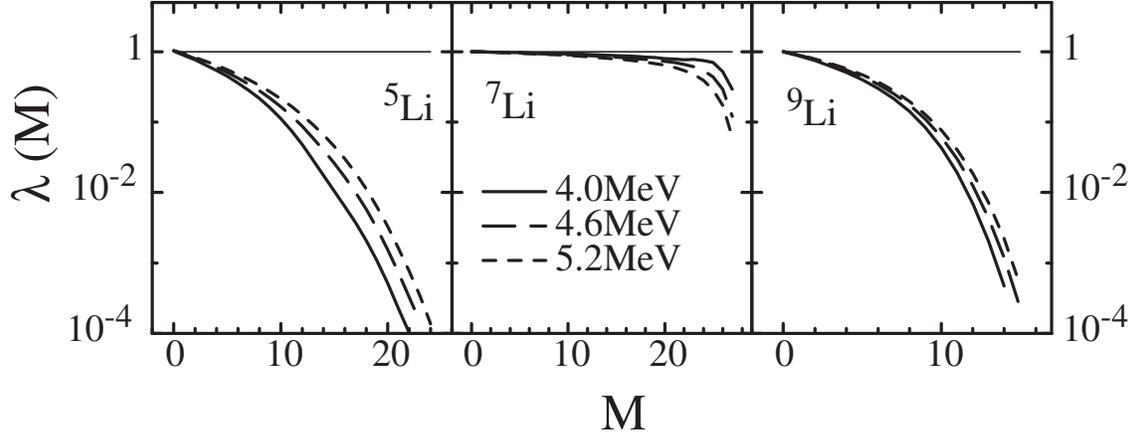}             %{m8-00b.eps}
\caption{The Poisson test parameter for multiplicity distribution of single clusters, Z=3, A=5,7,9, at three different temperatures $T=4.0, 4.6$, and $5.2$ MeV. The system is $(A_{sys}=200,Z_{sys}=86)$ and phase transition temperature is $T_b=4.6$ MeV for this system. The line at $\lambda(M)=1$ is for exact Poisson distribution.} 
 \label{fig:comp2:mdist-single-pstest-Trend}
\end{figure}
 
The multiplicity distribution of a single isotope has similar gross features whether in two component system or one component system. The exponential small probability for high multiplicity and the almost perfect fit with a Poisson distribution are also true for the two component model, see Fig.\ref{fig:comp2:mdist-single-Li}. Similar as we discussed before, such features shows that the production of a single isotope is just a small fraction of the total phase space. We can also use the Poisson test parameter to gauge the goodness of Poisson fit for the multiplicity distributions in two component system, and the results are shown in Fig.\ref{fig:comp2:mdist-single-pstest-Trend}. The multiplicity distributions of single isotopes showed interesting isospin dependence, which is strongly related to the isospin asymmetry of the whole system. The fragmentation source has isospin asymmetry of $\delta=(A-2Z)/A=0.14$, the isospin asymmetry of the selected isotopes are $\delta(^{5,7,9}Li)=-0.2,0.143,0.33$ respectively. While the isospin of isotope $^7Li$ is most close to the whole system, the Poisson test parameter for $^7Li$ is much closer to unity than any other isotopes. The Poisson test parameters for $^6Li$  lie between the values for $^5Li$ and $^7Li$, and is not plotted for clarity. Similarly, Poisson test parameters for $^8Li$ lie between the values for $^7Li$ and $^9Li$.

In the language of constrained partition problem, clusters with isospin asymmetry similar to the whole system, $^7Li$ in the current case, will feel the constraint of isospin much later than other clusters.
For the production of $^5Li$, multiple $^5Li$ in the system forms an extreme proton-rich subsystem, which is far away from our initial neutron-rich system. To reach our expected isospin asymmetry in the whole system, we have to compensate this "neutron deficiency" with several extreme neutron-rich isotopes to get the total isospin asymmetry right. And we know that clusters with extreme isospin asymmetry has much lower formation probability as compared to other similar sized clusters.
This means that $^5Li$ will feel the isospin constraint (or boundary condition) at quite low multiplicity. 
We may also think of produced isotope as building blocks from which we form the whole system. To form a system with isospin asymmetry $\delta=0.14$, it is of course easy to use building blocks with similar isospin asymmetry, such as $^7Li$. Then $^7Li$ will be more freely used than other isotopes with quite disparate isospin asymmetry.

The multiplicity distributions at different temperatures also show strong isospin dependence. As shown in Fig.\ref{fig:comp2:mdist-single-pstest-Trend}, the Poisson test parameter increases with increasing temperature for both $^5Li$ and $^9Li$, but the opposite is true for $^7Li$. The general trend of $^7Li$ is in fact the same as the result for single component system, compare to Fig.\ref{fig:comp1:mdist-singles-pstest}, and should not be a surprise. Then why does all other isotopes will have a different temperature dependence for the Poisson test parameter? In the scale as used for $^5Li$, we see that the lines for $^7Li$ is practically unity at all three temperatures. In this sense, we should ask why the multiplicity distribution of $^5Li$ looks more like that of $^7Li$, as temperature increases. Generally, isospin asymmetry in the total system is a small perturbation parameter, so that the total partition function may be expanded in isospin asymmetry, where the coefficient in front of isospin asymmetry will behave like an effective control parameter. The effect of such control parameter, in general, will get reduced at higher temperatures, so that the isospin difference between $^5Li$ and $^7Li$ became less important in the partition problem.

In the one component model, there is a critical change from super-Poissonion to sub-Poissonion distributions when the temperature crosses the phase transition temperature as shown in Fig.\ref{fig:comp1:mdist-singles-pstest}. But as shown in Fig.\ref{fig:comp2:mdist-single-pstest-Trend}, such a critical change is not seen in the range of temperatures we considered. If such a critical behavior does exist, the corresponding temperature should be much lower than the temperatures shown here. This is in agreement with conclusions in Sects.\ref{sect:gatedyield} and \ref{sect:varianceandmean}.

%%% decay and incomplete converage
\subsection{decay and feeding}

In the fragmentation model, clusters are produced "hot", and could decay into other particles. Such decay process will change the yield of clusters dramatically \cite{Botvina:1987jp,Souza:2000,Liu:2002tp}. For the distribution of given clusters $k$, it has two effects: the cluster $k$ could decay into other clusters, which we shall term as "decay" for now; and the decay of other clusters into cluster $k$, which shall be termed as "feeding". If the one of the interested cluster decays, we may find a multiplicity M event be identified as a multiplicity M-1 event. When feeding happens, we will misidentify a multiplicity M event as a multiplicity M+1 event. Note that experimentally, the incomplete coverage of the solid angles will produce the same effect as the decay process.

\begin{figure}[tbph!]
\center
\includegraphics[scale=0.6]{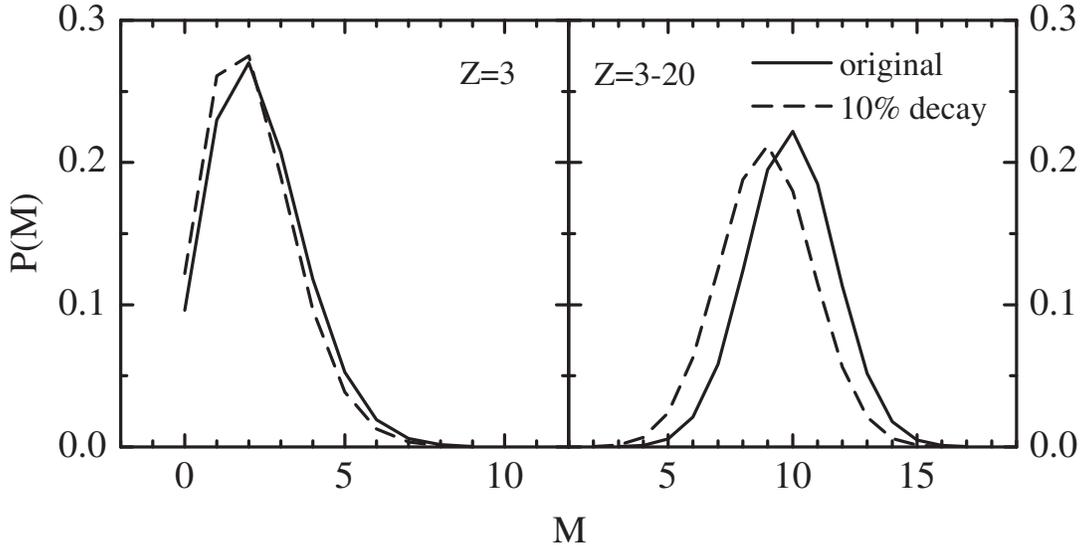}     %{m9-00b.eps}
\caption{The effect of simple decay on the multiplicity distribution of IMFs. We have plotted the before and after decay multiplicity distribution of single element (Z=3), and IMFs ($3 \leq Z \leq 20$) in the left and right panel respectively. The system is $(A_{sys}=200,Z_{sys}=86)$ at $T=5.2$ MeV. The decay probability is assumed to be $10\%$.} 
 \label{fig:comp2:mdist-simpledecay}
\end{figure}

Let us first consider the simple decay without any feeding process. For distributions with a mean multiplicity $\langle M \rangle$ much less than $1$, the decay will keep the general shape of the distribution, that is, $P(M)$ for each higher $M$ is order of magnitude smaller than the previous one. So that the difference from decay process is not readily seen in a simple multiplicity distribution plot. But for distributions with a higher mean multiplicity, $\langle M \rangle \gg 1$, the decay process obviously changes the shape of the distribution considerably, as shown in Fig.\ref{fig:comp2:mdist-simpledecay}.

\begin{figure}[tbph!]
\center
\includegraphics[scale=0.6] {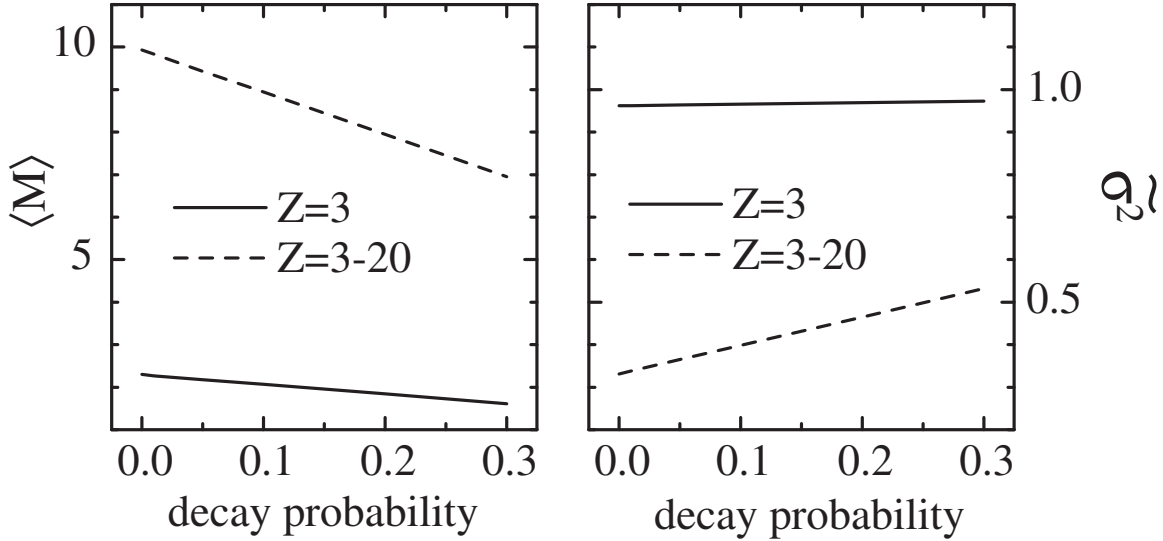}       %{m10-00b.eps}
\caption{The mean and variance of multiplicity distributions change as a function of decay probability in two comp model. Here we have use the same multiplicity distribution of single element (Z=3), and IMFs ($3 \leq Z \leq 20$) as shown in Fig.\ref{fig:comp2:mdist-simpledecay}, but the resulting linear relation is universal. The mean $\langle M \rangle$ is defined in Eq.(\ref{eq:sect5:eq1}), and scaled variance $\widetilde{\sigma^2}$ is defined in Eq.(\ref{eq:sect5:eq3}).} 
 \label{fig:comp2:mdist-mean-variance-simpledecay}
\end{figure} 

It is easiest to look at the changes of the mean multiplicity $\langle M \rangle$ and the variance $\widetilde{\sigma^2} = \sigma^2 /\langle M \rangle$ of the distribution, and the results are shown in Fig. \ref{fig:comp2:mdist-mean-variance-simpledecay}. The exact definitions for mean and variance are given in Sect.\ref{sect:varianceandmean}, see Eqs.\ref{eq:sect5:eq1} and \ref{eq:sect5:eq3}.
Since typical cluster decay probability for the produced "hot" IMF clusters is on the order of $0.10$, we are assured that the mean value and the variance for the multiplicity distribution are not changed too much. But in the case of incomplete solid angle coverage, we assume the "effective" decay probability could be as large as 30\%. 

\begin{figure}[tbph!]
\center
\includegraphics[scale=0.4]{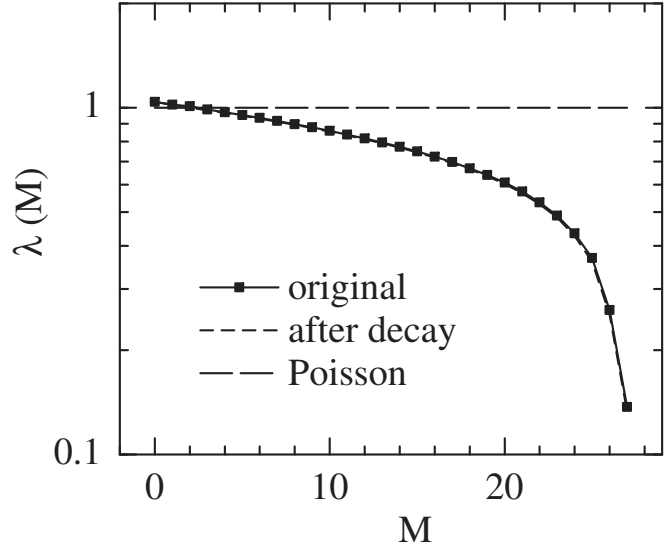}         %{m11-00.eps}
\caption{The Poisson test parameter $\lambda(M)$ for multiplicity distribution of a single element, Z=3, as a function of multiplicity $M$. The effect of simple decay will not alter the Poisson test parameter. The system is $(A_{sys}=200,Z_{sys}=86)$ at $T=5.2$ MeV. } 
 \label{fig:comp2:mdist-pstest-simpledecay}
\end{figure}

As shown in Fig.\ref{fig:comp2:mdist-simpledecay}, the details of the multiplicity distributions are indeed significantly changed by the simple decay process, especially in the case of distributions of high mean multiplicity events. We want to see if the Poisson test parameter is changed by such decay process or not.
For this, we have plotted the Poisson test parameter after a simple decay process in Fig.\ref{fig:comp2:mdist-pstest-simpledecay}. Interestingly enough, the simple decay process does not change the general systematics of the multiplicity distribution in the Poisson test parameter plot. This suggests that Poisson test parameter is a good observable for comparing the current fragmentation model with the experiments.

%%%2 comp, range IMF 

\begin{figure}[tbph!]
\center
\includegraphics[scale=0.6]{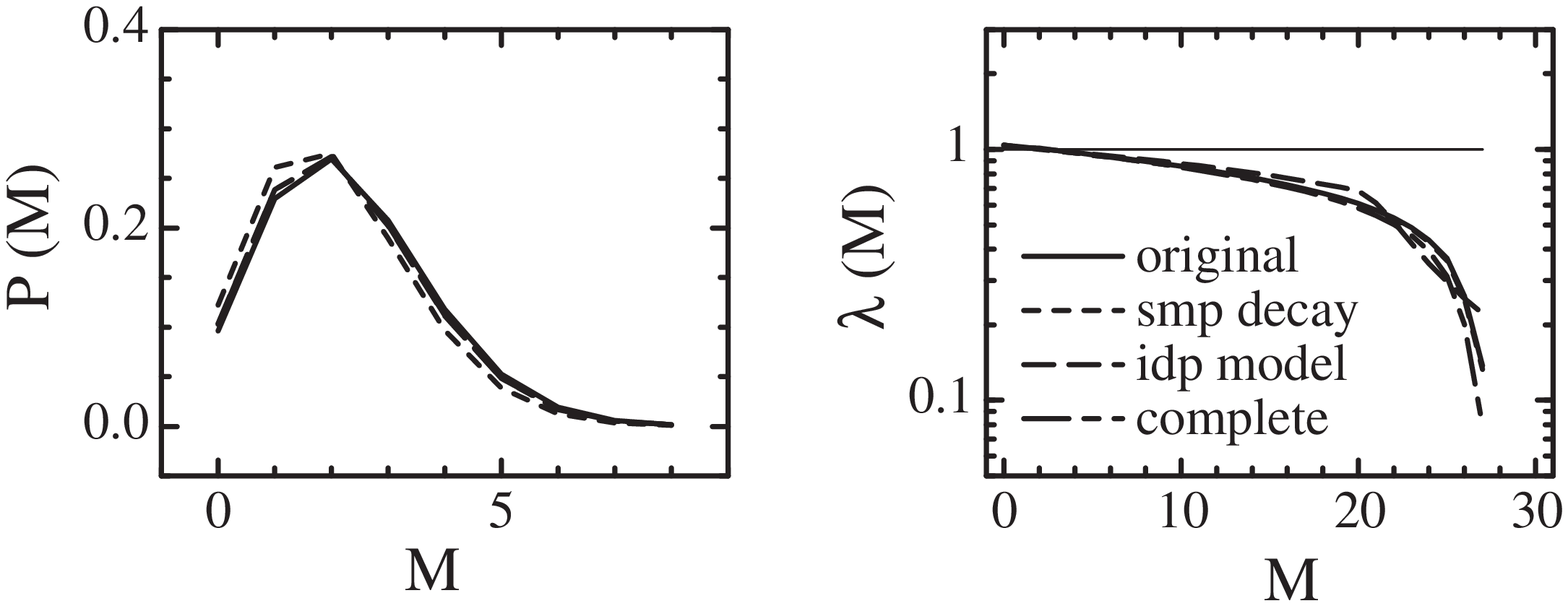}       %{m12-01.eps}
\caption{The effects of decay and feeding on the multiplicity distribution of element $Z=3$. The left panel shows the multiplicity distributions in linear scale, while the right panel shows the Poisson test parameter for the different multiplicity distributions. The line labelled "original" is the distribution without any consideration of either decay or feeding; the line labelled "smp decay" only includes the effect of decay process; the line labelled "complete" is the exact result which includes both decay and feeding; the line labelled "idp model" assumes the production of clusters are independent of each other, see Eq.(\ref{eq:sect3:eq2}). In general, inclusion of feeding process reduces the effect of decay and produces results more closer to "original" distribution. Such decay and feeding process will not change the systematics of the Poisson test parameter. Also not quite easy to see in the left panel, the "idp model" yields values that are up to three significant digits as good as the "complete" calculation in the low multiplicities $M<10$.} 
 \label{fig:comp2:mdist-completedecayfeeding}
\end{figure}    

%%% decay+feeding
The feeding process just has the opposite effect to the decay process. In general, it will reduce the changes of the multiplicity distribution caused by the decay process. The decay probability of cluster $k+1 \rightarrow k$ is generally different than the decay probability of cluster $k \rightarrow k-1$. For simplicity of discussion, we will assume that the decay probability of clusters $k+1$ and $k$ are the same in the following analysis. 

The complete solution to the decay and feeding problem requires the tagging of separately the decay group and the feeding group, and is indeed more complicated than simple decay problem. Fortunately, such problem is equivalent to a new partition problem as shown in Appendix \ref{appendix:decay}, and could be solved exactly.
Such a repartition method requires a redefinition of the cluster group and the cluster formation probability, but the partition part is the same as the tagged multiplicity distribution problem. The results for the multiplicity distribution after the complete decay and feeding process is shown in Fig.\ref{fig:comp2:mdist-completedecayfeeding}.
As we can see, the Poisson test parameter is almost not changed by the decay and feeding process.

The independent production ("idp" in Fig.\ref{fig:comp2:mdist-completedecayfeeding}) model assumes the production of clusters are independent of each other. In more precise term, the probability of having $N_1$ clusters in the group $alpha$ and $N_2$ clusters in the group $beta$ is:
\begin{eqnarray}
P(\alpha,N_1; \beta, N_2) \approx P(\alpha,N_1) P(\beta,N_2) \,.      
  \label{eq:sect3:eq2}
\end{eqnarray}  
With this combined formation probability, we can employ the same simple decay method as shown in the first part of Appendix \ref{appendix:decay} and do not need to build a new partition in the system. The independent production approximation in fact works quite well for the case of Z=3 cluster group, matches well to that of the exact solution up to three significant digits in the low multiplicity region. Actually, such statistical independence between produced clusters are quite expected. Since the formation of a single cluster is only a small fraction of the total fragmentation channel, there is little correlation between the two type of clusters, unless each group contains far more than one clusters, in which case, the correlation could build up. The deviations of the independent production model from the exact result are easily seen in high multiplicity events as shown in the right panel of Fig.\ref{fig:comp2:mdist-completedecayfeeding}.

\section{multiplicity and system size}
\label{sect:systemsize}

%%% sysmatics of multiplicity distribution, as a function of system size

The multiplicity distributions as a function of total transverse energy are measured in experiments \cite{Moretto:1995,Beaulieu:1998a,Beaulieu:1998b}. As the transverse energy is roughly correlated with the centrality of the reaction, we may interpret the results as a system size dependence. But the reader is cautioned that when the transverse energy goes all the way down to zero, the IMF fragment production mechanism has changed from a central collision to a peripheral collision. To be consistent, we will only consider central to semi-central collision where intermediate sized clusters are profusely produced, and the production mechanism is better understood. The current thermodynamic model is, in any case, inapplicable at low temperatures where the fragment production is most likely sequential.

\begin{figure}[tbph!]
\center
\includegraphics[scale=0.4]{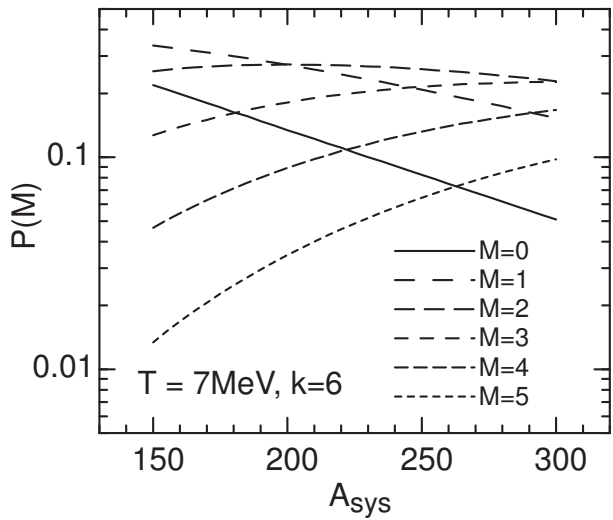}      %{m13-00b.eps}
\caption{Multiplicity distribution of a single cluster $k=6$ as a function of system size at fixed temperature $T=7$ MeV in one component model.} 
 \label{fig:comp1:mdist-singles-systemsize}
\end{figure}

\begin{figure}[tbph!]
\center
\includegraphics[scale=0.4]{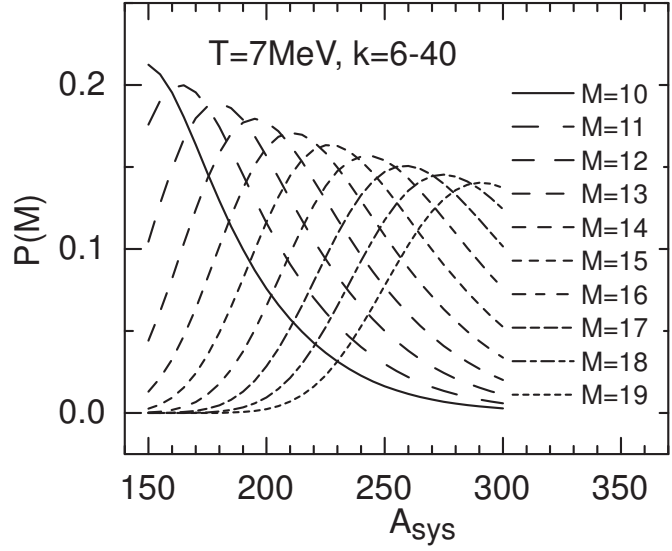}      %{m14-01b.eps}
\caption{Multiplicity distribution of a range of clusters $k=6-40$ are plotted as a function of system size at fixed temperature $T=7$ MeV in one component model. Only multiplicities in the range of $10 \leq M < 20$ are plotted here.} 
 \label{fig:comp1:mdist-IMF-systemsize}
\end{figure}

%%%size dependence   one comp
In the one component model, the multiplicity distribution of a single cluster $k=6$ is shown as a function of system size in Fig.\ref{fig:comp1:mdist-singles-systemsize}.
As the mean multiplicity for a single IMF is quite small, only the first few of the P(M)'s are significantly above $10^{-2}$ at $T=7$ MeV. At fixed temperature, the crossings of different P(M)'s are strongly correlated with the total mass of the fragmentation source. Such result is consistent with the overall features of multiplicity distribution for single element in \cite{Beaulieu:1998a}. We can expect each P(M)'s will rise, reach a maximum and then fall as system size increases. Such a trend is more clear in Fig.\ref{fig:comp1:mdist-IMF-systemsize}. Here we will refrain from comparing results from one component model with the experimental results because the temperature correspondence are quite different in one and two component systems.

The multiplicity distribution for a range of IMFs $6 \leq k \leq 40$ is shown in Fig.\ref{fig:comp1:mdist-IMF-systemsize}. As with the multiplicity distribution for single clusters, the crossing of different multiplicity lines for a range of clusters are also correlated with the system size. As evident in Fig.\ref{fig:comp1:mdist-IMF-systemsize}, the different multiplicity line has very similar shape for $M=10$ all the way up to $M=19$. The peak position of different multiplicity line shifts by a constant when the system size increase, and the peak value slightly decreases.

%% size dependence   2 comp

\begin{figure}[tbph!]
\center
\includegraphics[scale=0.4]{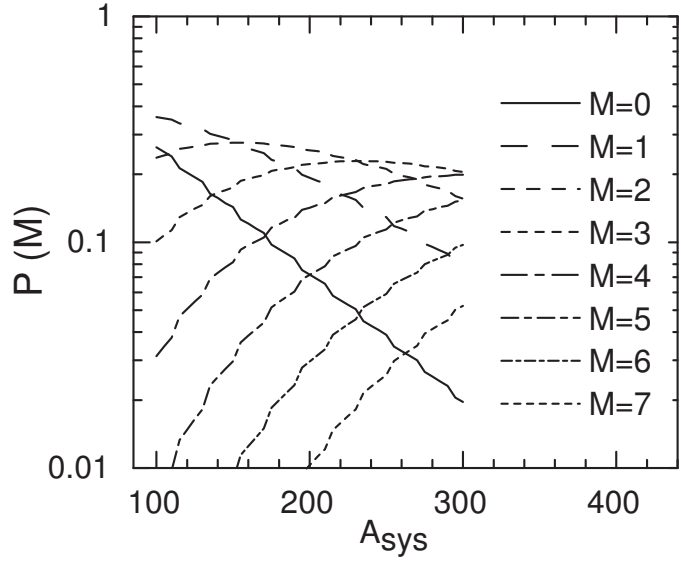}      %{m15-00b.eps}
\caption{Multiplicity distribution of $Z=3$ is plotted as a function of system size at fixed temperature of 5.2 MeV in a two component model. Charge number of the fragmentation source is chosen so that the total system isospin asymmetry is as close to $\delta=0.1$ as possible.} 
 \label{fig:comp2:mdist-singles-systemsize}
\end{figure}

\begin{figure}[tbph!]
\center
\includegraphics[scale=0.4]{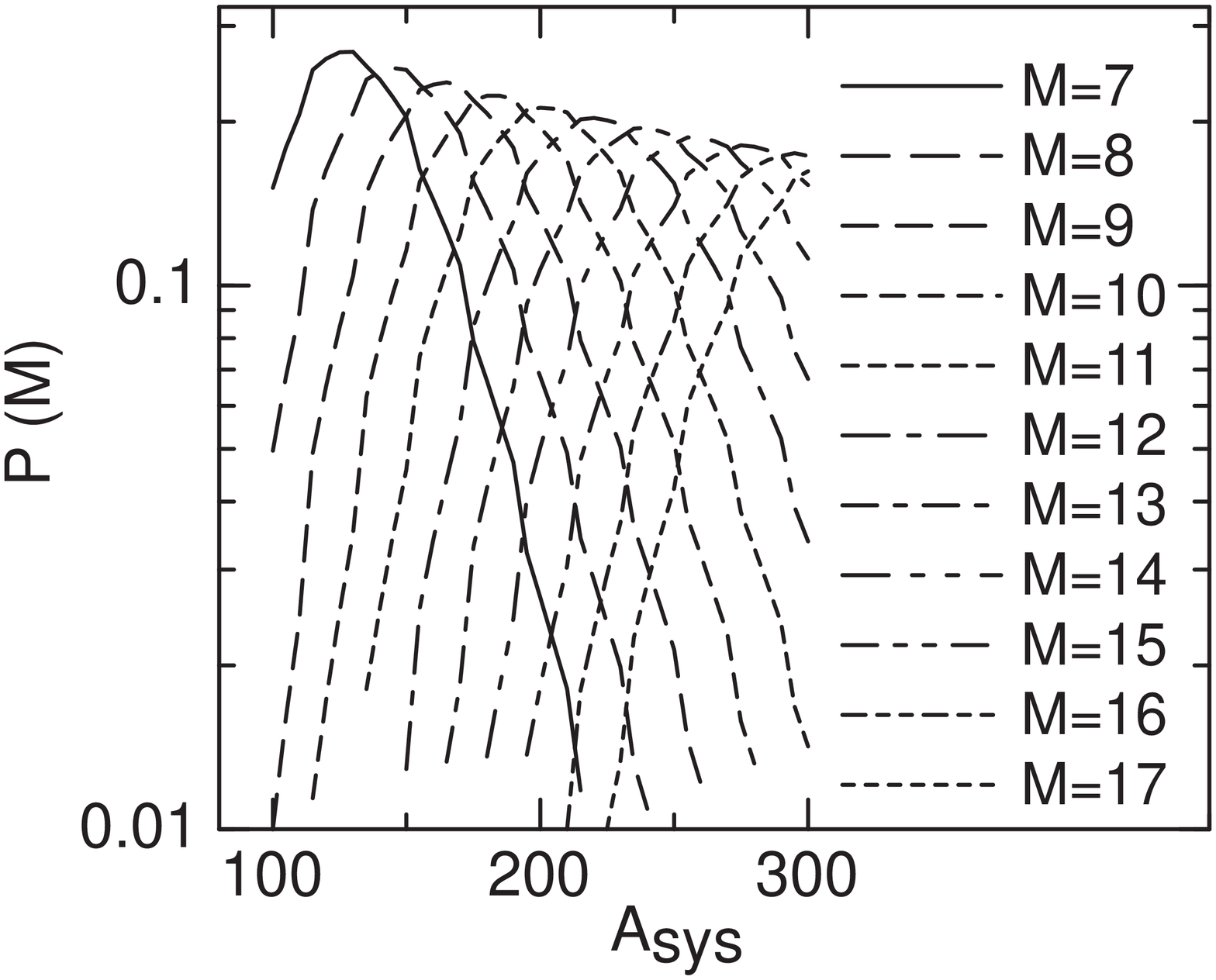}       %{m15+00b.eps}
\caption{Multiplicity distribution of IMFs $3 \leq Z \leq 20$, is plotted as a function of system size at fixed temperature of 5.2 MeV in a two component model. The Charge number of the source is chosen so that the total system isospin asymmetry is s close to $\delta=0.1$ as possible. Only lines with multiplicity in the range $7 \leq M \leq 17$ are plotted.} 
 \label{fig:comp2:mdist-IMF-systemsize}
\end{figure}

Since a temperature around $T=4-6$ MeV is often cited in intermediate energy heavy-ion reactions \cite{Jaqaman:1983,Lynch:1987,Tsang:1997,Schwarz:1993,Pochodzalla:1995xy,D'Agostino:1999is,Milazzo:2002kr,D'Agostino:2003pw}, we have plotted the multiplicity distributions of a single element and a range of IMFs as a function of system in Figs.\ref{fig:comp2:mdist-singles-systemsize} and \ref{fig:comp2:mdist-IMF-systemsize} at a fixed temperature of $T=5.2$ MeV. In producing both figures, we try to keep the system isospin asymmetry close to $\delta=0.1$, but as system size and charge are integers, we can only select the closest integer number. As can be seen from these two figures, the variation of isospin does not change much of the systematics related to system size. Other than the variation of isospin, we find the results for multiplicity distribution in two component system and one component system have similar systematics.

At this stage, we can try to compare the systematics of multiplicity distributions with that from experiments. Specifically, Beaulieu et al. presented the systematics of multiplicity distribution of $Z=3$ for the reaction $Xe+Au$ at $50$ AMeV in Fig.4 in \cite{Beaulieu:1998a}. The transverse energy range plotted there is quite large $0 \leq E_T \leq 1000 MeV$. But since the low transverse energy data are most probably produced in peripheral reactions where the cluster production mechanism may be of a sequential nature, we will only concentrate on the more central collisions where the transverse energy is in the range of $500 MeV \leq E_T \leq 1000 MeV$. The total system size could be assumed to be proportional to the total transverse energy, while the system temperature is roughly constant, thus we estimate the system size to be $120 \leq A_{sys} \leq 240$. This assumption is of course too simplified. There may also be a change of temperature when the transverse energy $E_T$ changes. 
For simplicity, we also assumed the total system isospin asymmetry is roughly $\delta=0.10$.

As can be seen from Fig.4 in \cite{Beaulieu:1998a}, the mean multiplicity of Li is considerably lower than that in Fig.\ref{fig:comp2:mdist-singles-systemsize}. To reach lower mean multiplicity, we either have to choose a smaller system or a lower temperature. On top of these two uncertainties, we do have another freedom to change the decay probability for Li element to simulate the incomplete coverage of emitting angles. So we choose to fix the system size as we have estimated, and try to fit the experimental result with only variation of temperature and decay probability. Additionally, a large decay probability and a lower temperature have similar effect in reducing the mean multiplicity, with the temperature effect being the more prominent of the two. So we may constrain ourselves to a high temperature and lower decay probability. Since in fragmentation models, a temperature of around 5 MeV is often used \cite{D'Agostino:1999is,Tsang:1997}, we will try to find a sensible decay rate while keeping temperature as close to 5 MeV as possible. With the current published data, we have found no reference to get an estimate on the effective decay probability, we will just assume it is between 15\% and 30\%. Since this decay probability also includes the incomplete coverage of the detector system, this range is not too far away from reality.

\begin{figure}[tbph!]
\center
\includegraphics[scale=0.4]{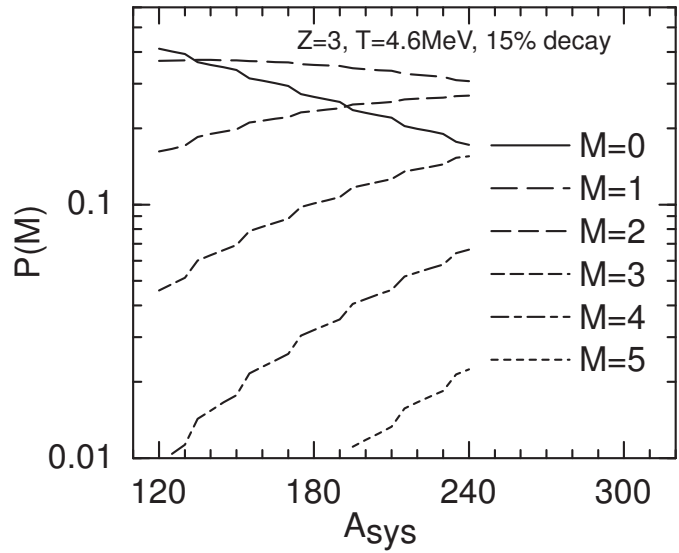}        % {m16-00.eps}
\caption{Multiplicity distribution, for single element $Z=3$, is plotted as a function of system size in a two component model. We have assumed 15\% decay probability for all Li isotopes and the temperature of the fragmentation source is fixed at T=4.6 MeV. Such result roughly corresponds to the multiplicity distribution in the range $500 MeV \leq E_T \leq 1000 MeV$ at the left panel of fig.4 in \cite{Beaulieu:1998a}.} 
 \label{fig:comp2:mdist-IMF-systemsize-decay-single}
\end{figure}

\begin{figure}[tbph!]
\center
\includegraphics[scale=0.4]{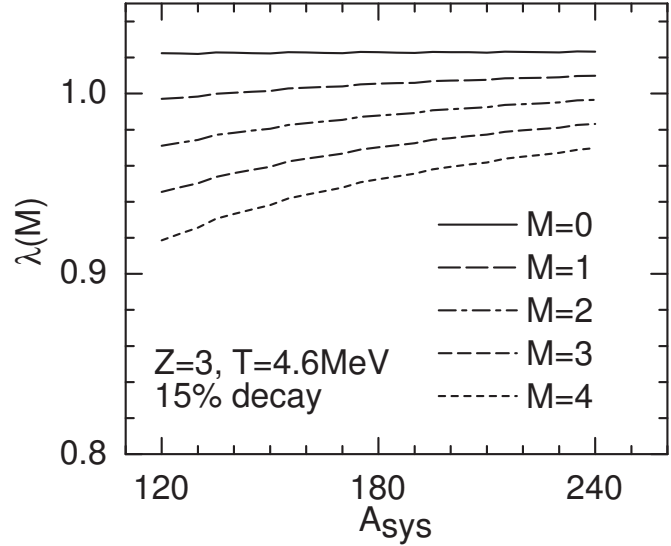}     %{m16+00.eps}
\caption{The Poisson test parameter, for single element $Z=3$, is plotted as a function of system size. We have assumed 15\% decay probability for all Li isotopes and the temperature of the fragmentation source is fixed at T=4.6 MeV. This result roughly corresponds to the right panel of fig.4 in \cite{Beaulieu:1998a}, with the range $500 MeV \leq E_T \leq 1000 MeV$. Note, our definition of Poisson test parameter $\lambda (M)$ scaled out the obvious factor of mean multiplicity, see Eq.(\ref{eq:sect3:eq1}).} 
 \label{fig:comp2:mdist-IMF-systemsize-decay-single-ps}
\end{figure}

We have plotted the multiplicity distribution of Li as a function of system size in Fig.\ref{fig:comp2:mdist-IMF-systemsize-decay-single}. A temperature of $T=4.6$ MeV and a decay probability of $15\%$ seem to give a good fit to the data in Fig.4 in \cite{Beaulieu:1998a}. Combinations of either a slightly lower temperature and lower decay probability or slightly higher temperature and higher decay probability also gives similar results. 

The Poisson test parameter for the same distribution in Fig.\ref{fig:comp2:mdist-IMF-systemsize-decay-single} is plotted in Fig.\ref{fig:comp2:mdist-IMF-systemsize-decay-single-ps}. The this result could be compared with the right panel of Fig.4 in \cite{Beaulieu:1998a}. Note in the definition of Poisson test parameter Eq.(\ref{eq:sect3:eq1}), we have scaled out the mean multiplicity. The results seem to agree very well with the systematics in \cite{Beaulieu:1998a}.

The multiplicity distribution of a range of IMFs, $3 \leq Z \leq 20$, is also presented in \cite{Moretto:1995}. In the reaction of $Ar+Au$ at $110$ AMeV, multiplicity distributions of IMFs are measured in the transverse energy range $0 \leq E_T \leq 1100 MeV$. As before, we will try to reproduce the multiplicity distributions in the more central collisions with $550 \leq E_T \leq 1100 MeV$, and the system size is estimated as $110 \leq A_{sys} \leq 220$. Again, we have assumed the total system isospin asymmetry to be around $\delta=0.10$.

\begin{figure}[tbph!]
\center
\includegraphics[scale=0.4]{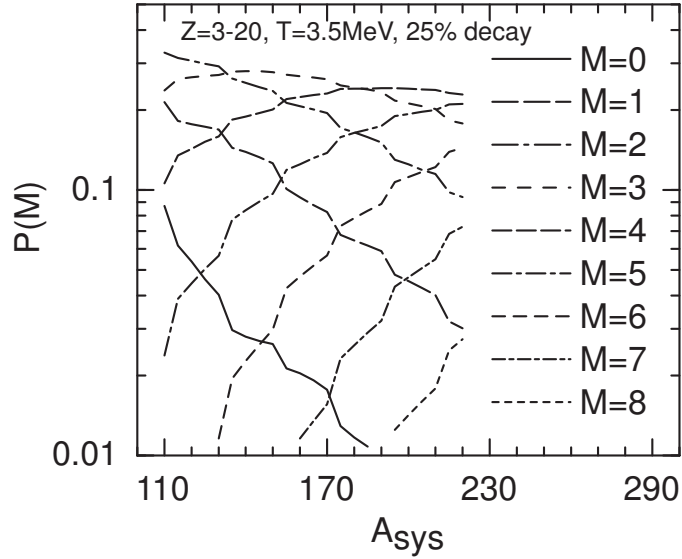}   %{m17-00.eps}
\caption{Multiplicity distribution, for a range of IMFs $3 \leq Z \leq 20$, is plotted as a function of system size for two component system. We have assumed 25\% decay probability for all Li isotopes and the temperature of the fragmentation source is fixed at T=3.5 MeV. This result roughly corresponds to the range of $550 MeV \leq E_T \leq 1100 MeV$ in the top panel of Fig.3 in \cite{Moretto:1995}.} 
 \label{fig:comp2:mdist-IMF-systemsize-25decay1}
\end{figure}

As shown in Fig.\ref{fig:comp2:mdist-IMF-systemsize-25decay1}, if we assume a temperature of $3.5$ MeV and a decay probability of $25\%$, the systematics of the multiplicity distribution reasonably produces the result in \cite{Moretto:1995}. Here the temperature of the source is significantly lower than the previous fitted $4.6$ MeV for $Xe+Au$ reaction. 

\begin{figure}[tbph!]
\center
\includegraphics[scale=0.4]{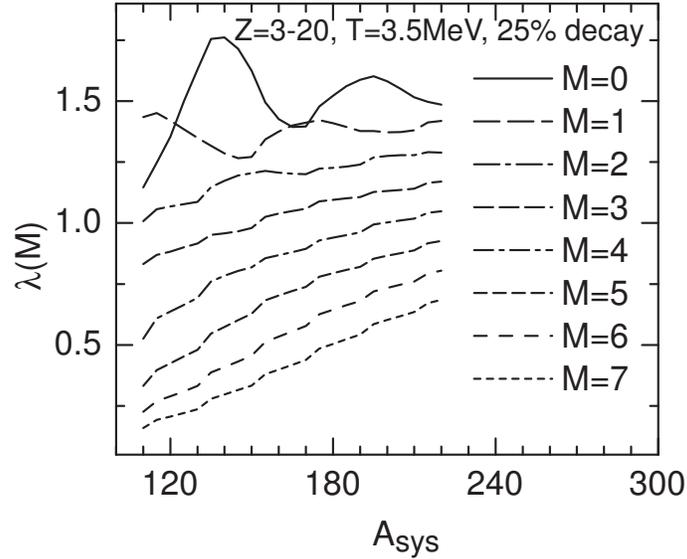}     %{m17+00.eps}
\caption{The Poisson test parameter, for a range of IMFs $3 \leq Z \leq 20$, is plotted as a function of system size for two component system. The IMF selection and $E_T$ correspondence are the  same as in Fig.\ref{fig:comp2:mdist-IMF-systemsize-25decay1}. We have assumed 25\% decay probability for all Li isotopes and the temperature of the fragmentation source is fixed at T=3.5 MeV.} 
 \label{fig:comp2:mdist-IMF-systemsize-25decay2}
\end{figure} 

In Fig.\ref{fig:comp2:mdist-IMF-systemsize-25decay2}, we have plotted the Poisson test parameter for the same multiplicity distributions in Fig.\ref{fig:comp2:mdist-IMF-systemsize-25decay1}. This result can be compared with that in Fig.\ref{fig:comp2:mdist-IMF-systemsize-decay-single-ps}. In the single element case, the Poisson parameters are always quite close to unity; while its magnitude changes considerably for a range of IMFs. This is also a manifestation of the correlations between IMF clusters, which will give rise to stronger signals for a large range of IMFs than for a single IMF (see also discussion in Sect. \ref{sect:varianceandmean}). The line for $\lambda(M=0)$ shows oscillations as a function of system size. This oscillations may be related to the "long range correlations" in the cluster production, where particular patterns of cluster formation matches well with the total system size and thus get enhanced at the optimum system size, and may get suppressed when mismatch happens. We will see more of this matching effect in Sect. \ref{sect:gatedyield}.   

\section{Multiplicity gated yield ratio}
\label{sect:gatedyield}

The most elementary measurements are inclusive cross sections. These are given by the yield $\langle n_k \rangle$, see Eq.(\ref{eq:sect2:eq4}). The isotope yield ratios are often used in heavy-ion reactions because some experimental uncertainties are cancelled in the ratios. Examples include single isotope yield ratios, double isotope yield ratios, and ratios of same isotope but from different reactions \cite{D'Agostino:1999is,D'Agostino:2003pw,Milazzo:2002kr,Tsang:1997,Souza:2000,Pochodzalla:1995xy,Liu:2002tp,Tsang:2001dk}.  It is also possible to make more complicated measurements: measure $\langle n_k \rangle$ subject to the condition that the event has a given IMF multiplicity $M$. We can define then a ratio: 
\begin{eqnarray}
R_k(M) = \frac{ \langle n_k(M) \rangle }{\langle n_k \rangle}
\,,       
  \label{eq:sect4:eq1}
\end{eqnarray}
where $\langle n_k(M) \rangle$ is the yield when the IMF multiplicity is $M$ and $\langle n_k \rangle$ is the usual (ungated) yield. 
 
%%%M gate yield ratio      one comp not done yet
\begin{figure}[tbph!]
\center
\includegraphics[scale=0.75]{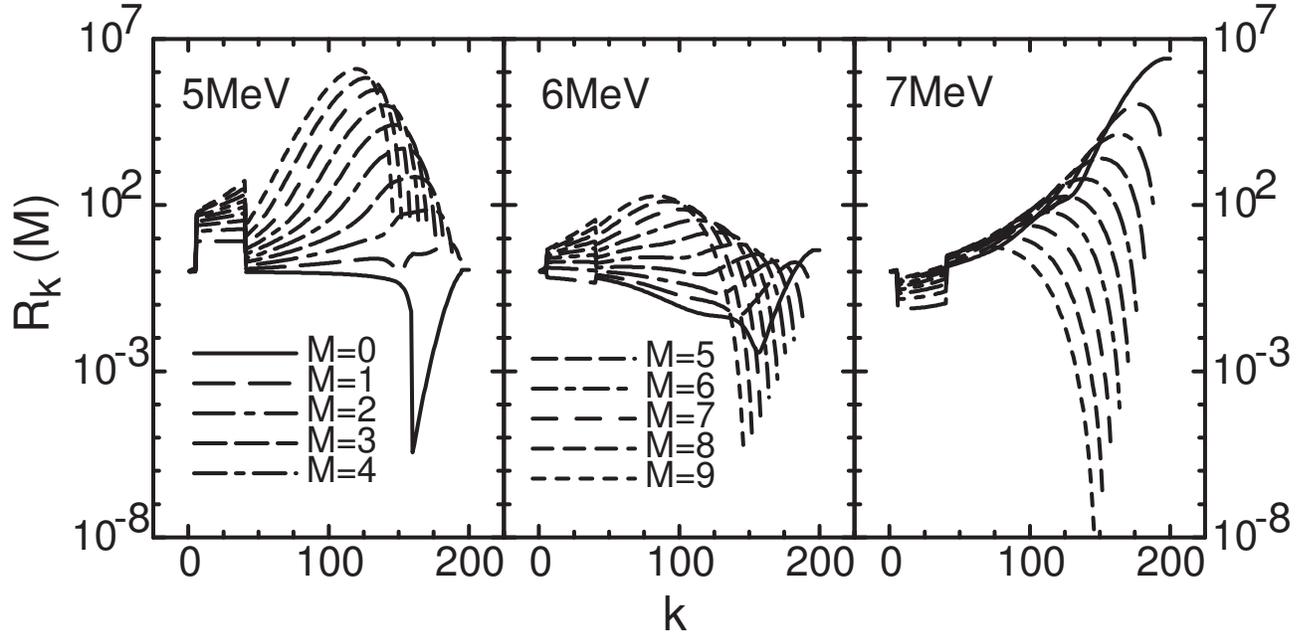}    %{m18-01-b02.eps}
\caption{The multiplicity gated to ungated particle yield ratio $R_k(M)$, for a single component system at around the phase transition temperature region. The three panels from left to right are for the same system of $A_{sys}=200$ particles at different temperatures $T=5, 6$ and $7$ MeV, while the phase transition temperature $T_b=6.5$MeV. For IMF in the range of $6 \leq k \leq 40$, the mean multiplicities at the three temperatures are $0.12, 1.55$, and $12.72$ respectively.} 
 \label{fig:comp1:IMF-gated-yield-ratio1}
\end{figure}

\begin{figure}[tbph!]
\center
\includegraphics[scale=0.5]{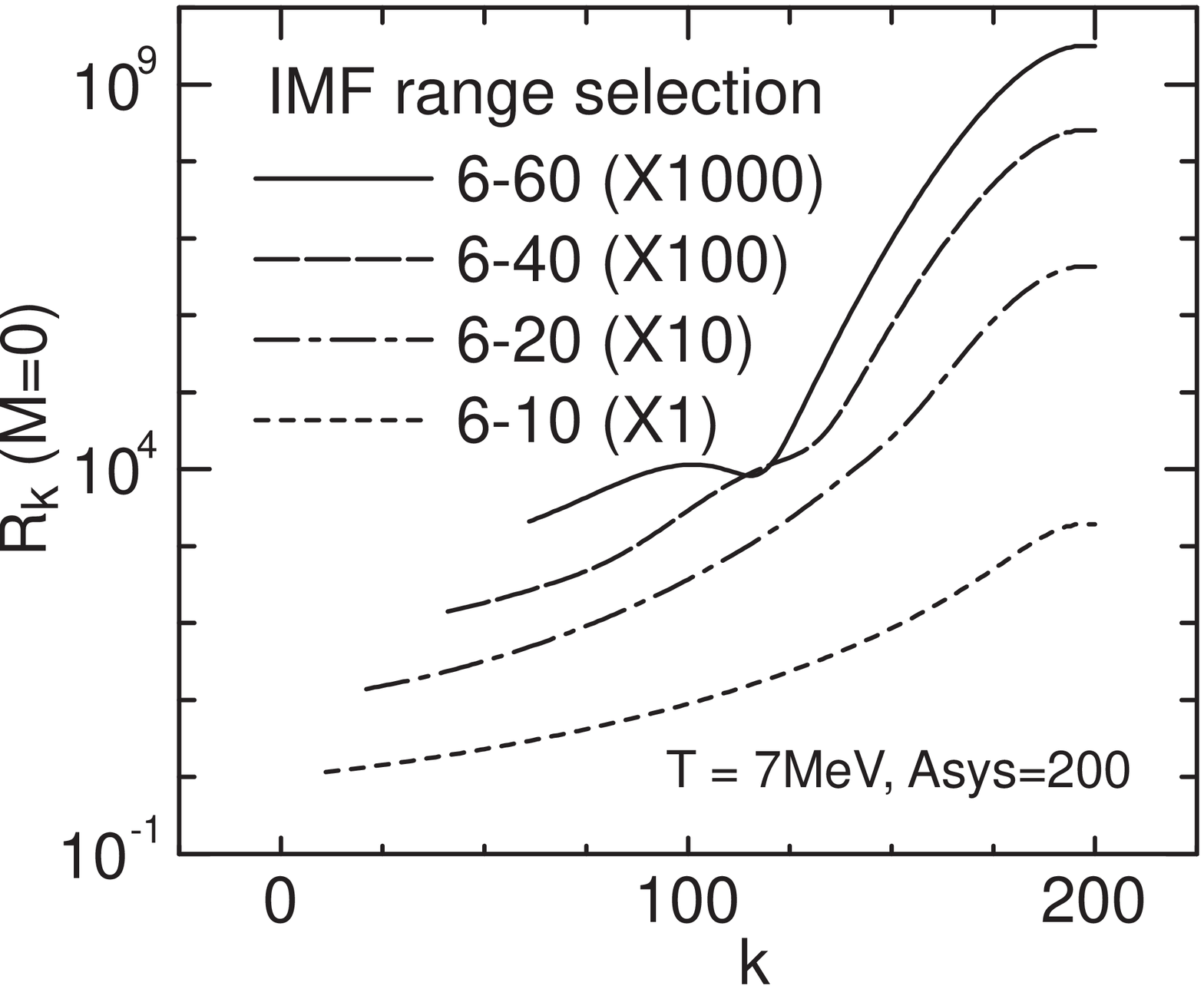}  %{m18+01-b01.eps}
\caption{The multiplicity gated to ungated element yield ratio $R_k(M)$, for a single component system, with $A_{sys}=200$ and $T=7MeV$.} 
 \label{fig:comp1:IMF-gated-yield-ratio2}
\end{figure}

In general, if we fix the IMF multiplicity to be lower than the mean IMF multiplicity, then the IMF production is artificially suppressed. If any other clusters are positively correlated with the IMF clusters, the yield of such clusters will also get suppressed. On the other hand, if the selected multiplicity is higher than the mean multiplicity, the IMF clusters get enhanced, and so will the positively correlated clusters.

The IMF multiplicity gated and ungated yield ratio $R_k(M)$ for a system of 200 particles at three different temperatures are plotted in Fig.\ref{fig:comp1:IMF-gated-yield-ratio1}. The three panels from left to right are results at temperatures well below phase transition temperature $T=5 MeV$, close to phase transition temperature $T=6$ MeV, and above phase transition temperature $T=7 MeV$. The corresponding mean multiplicity for the systems from left to right are $\langle M \rangle = 0.12, 1.55$, and $12.72 $ respectively. We can see a change of shape for different multiplicity lines when temperature changes from below to above phase transition temperature. Such a shape change is an exhibition of the critical behavior in the single component system.

Aside from the obvious shape change in Fig. \ref{fig:comp1:IMF-gated-yield-ratio1}, we can also find the correlation between IMF clusters and heavier mass clusters. At 5 MeV, most of the heavier mass clusters are positively correlated with the IMFs when $M>0$, and the line for $M=0$ is strongly suppressed; while at $7$ MeV, each of the $P(M)$'s has an positively correlated region and a strongly reduced correlation at heavier masses. The only exception is the multiplicity $M=0$ line, where only positive correlations is observed.

As we have indicated before, the selection of a range of IMF clusters will typically enhance the signals. Here we may test the enhancement by changing the selection for IMF clusters and see if there is any change in the pattern of yield ratio. We varied the IMF range and plotted the lines for yield ratio in Fig.\ref{fig:comp1:IMF-gated-yield-ratio2}. The multiplicity $M=0$ lines seem to be specially interesting, aside from simple change of magnitude in the yield ratio, certain oscillating structure shows up when the range of IMF is large enough. The appearance of oscillating structure is due to the "long range multiparticle correlations" between heavier clusters and will be elucidated in more detail here.  

In the case of high temperature $T=7$ MeV$>T_b$, the system is predominately in the gas phase, and small clusters are profusely produced, and these small clusters are highly correlated. All other larger clusters will have negative correlation with the IMFs, and a select of $M=0$ for IMFs suppress the productions of IMFs and enhance the production of larger clusters. The additional oscillations in the cluster yield ratio, represent the multiply correlations between clusters of quite different size, and thus are of "long range" nature. For example, the line labelled "$6-60$" in Fig.\ref{fig:comp1:IMF-gated-yield-ratio2}, has a significant dip at around $k=115$ while the line "6-40" does not show a dip at the same fragment size. This dip must come from the exclusion of IMFs in the range $40 < k \leq 60$. This could happen when the source prefers to fragment into two large pieces, one of size 115 and another of size between 40 and 60, and also some small clusters. Since the probability of $M=0$ event is rare, it is not a surprise to see the exotic fragmentation process.   

%%%M gate yeild ratio    2 comp done
\begin{figure}[tbph!]
\center
\includegraphics[scale=0.4]{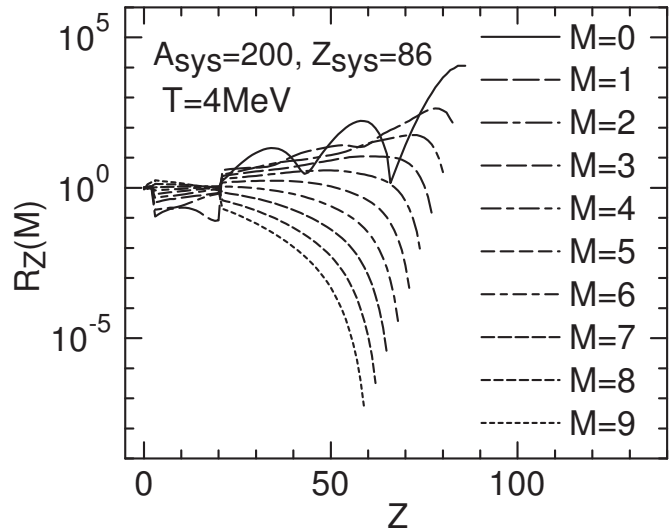}    %{m19-00b.eps}
\caption{The multiplicity gated to ungated charge yield ratio $R_Z(M)$ in a two component model. The total system is $A_{sys}=200$, $Z_{sys}=86$ at T=4 MeV. IMFs are defined as clusters with $3 \leq Z \leq 20$, and the mean IMF multiplicity is $\langle M \rangle =6.45$ in this system.} 
 \label{fig:comp2:IMF-gated-yield-ratio}
\end{figure}

\begin{figure}[tbph!]
\center
\includegraphics[scale=0.4]{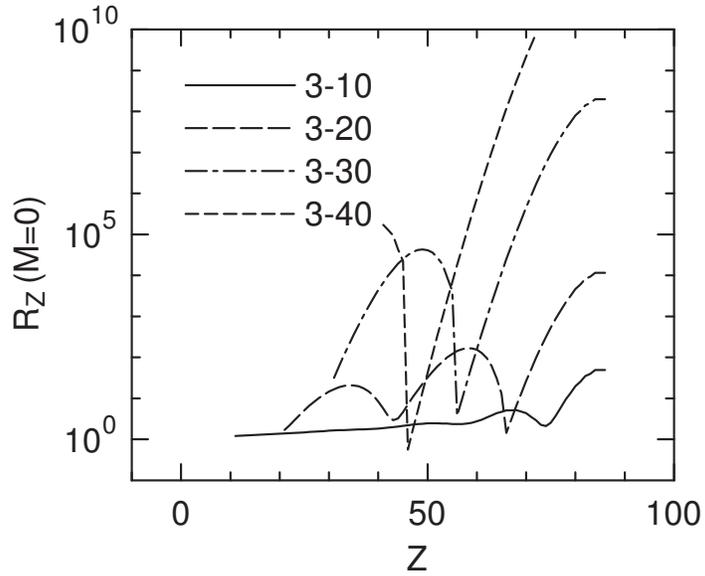}    %{m20-00b.eps}
\caption{Same at in Fig. \ref{fig:comp2:IMF-gated-yield-ratio}. The multiplicity gated to ungated charge yield ratio in a two component model where the IMF range selection is varied. The system is $A_{sys}=200$, $Z_{sys}=86$ at T=4 MeV. The different lines correspond to different IMF range higher cut-off from $Z \leq 10$ to $Z \leq 40$, while the lower cut-off is fixed at $Z \geq 3$.} 
 \label{fig:comp2:IMF-zero-yield-ratio}
\end{figure}

The gated to ungated charge yield ratio in a two component model shows similar features as in the one component model, but the change of shape does not show up at temperatures close to the phase transition temperature. Fig.\ref{fig:comp2:IMF-gated-yield-ratio} shows the typical yield ratio for a two component model, where the system size is taken as $A_{sys}=200, Z_{sys}=86$, and the temperature of the fragmentation source is well below phase transition temperature $T=4 MeV < T_b$. And if the shape change exists in the two component model, the corresponding temperature should be well below temperatures we are interested here. This result is consistent with the conclusion of Sect.\ref{sect:varianceandmean}.

Another interesting feature is the appearance of multiple oscillation in the line $M=0$ in Fig.\ref{fig:comp2:IMF-gated-yield-ratio}. The appearance of two dips at $Z=43, 66$ are pointing to two different exotic fragmentation channels, one is related to two clusters of size $Z\simeq20$ and $Z\simeq 66$, and another is related to three clusters of size $Z\simeq 43$ and two of $Z\simeq 20$. The significant charge difference between such clusters could be labelled as "long range", because most of the correlations we considered are between similar sized clusters, or "short ranged". It should be noted that the probability for these exotic fragmentation channels is quite small as compared with that of regular channels which also produce the heavy cluster $Z=66$ together with some IMF clusters. If we allow the production of small clusters in the range of $6\leq Z \leq 10$, the dip at $Z=66$ for will not show up, because the most of the channels producing a $Z=66$ are accompanied by several small clusters. 
On the other hand, the line for $M=1$ also shows oscillations and the phase of oscillations is just opposite to the line of $M=0$. The difference in the phase of the two lines is another manifestation of the strong correlation in the exotic fragmentation channels.

Fig.\ref{fig:comp2:IMF-zero-yield-ratio} shows the effect of different IMF range selections on the yield ratios for fixed IMF multiplicity $M=0$.
As with one component model, the enhanced yield of heavy elements are due to the negative correlations between IMFs and heavy clusters. The oscillation of the $M=0$ lines are more prominent in the two component model. The strong oscillation and the strong dip in the $M=0$ line suggest strong correlations between large cluster and small cluster. If we look at the largest charge value for the dip to happen $Z_{max}(dip)$, we find this values for different lines in Fig.\ref{fig:comp2:IMF-zero-yield-ratio} are of equal distance. The position of the dip is at $Z_{max}(dip)=74,66,56$,and $46$ respectively, and adding the cut-off value for the IMF range selection, we find the sum to be $84,86,86,86$. These exotic fragmentation channels are very similar to asymmetric binary fission.

\section{variance and phase transition}
\label{sect:varianceandmean}

The variance is one of the most simple cumulants for the distributions, and is often used as a synonym for fluctuation or correlation in the system. Higher order cumulants were also studied before \cite{Pratt:1999ht,Campi:1986,Campi:1995,Carruthers:1989}. Coincidentally, the variances for charge fluctuations \cite{Asakawa:2000,Jeon:2000}, transverse momentum fluctuations (see \cite{Mitchell:2004} and reference therein) are also proposed for relativistic heavy-ion collisions in the study of a possible phase transition to the quark-gluon plasma phase. Experimentally, the variance is easiest to measure. The variance is related to the correlations between particles, and the appearance of a maximum variance is sometimes interpreted as a critical process in intermediate energy heavy-ion collisions \cite{Pratt:1999ht,Campi:1986,Campi:1995}. The strong correlations between clusters, or fluctuations, are sometimes attributed to the phase transition in the fragmentation source \cite{Pratt:1999ht}. In this section, we will test the relation between correlations and phase transition and try to fit the fragment correlation results from experiments. 

The mean multiplicity and the variance are defined as
\begin{eqnarray}
\langle M \rangle &=& \sum_{M} M P(M)
 \, ,
 \nonumber \\
\sigma^2 &=&  \sum_{M} M^2 P(M) - \langle M \rangle^2   \,,         
  \label{eq:sect5:eq1}
\end{eqnarray}
where $M$ is the multiplicity of the interested particles. Note that from the definition of $P(M)$, the definitions above are complete equivalent to the usual definitions for mean multiplicity and variance:
\begin{eqnarray}
\langle M \rangle &=&  \langle \sum_{i \in \alpha}n_i \rangle 
 \, ,
 \nonumber \\
\sigma^2 &=&   \langle (\sum_{i \in \alpha}n_i)^2 \rangle -\langle \sum_{i \in \alpha}n_i \rangle ^2  \,,         
  \label{eq:sect5:eq1p}
\end{eqnarray} 
For Poisson distribution, the variance is exactly equal to the mean value, so it is instructive to use the scaled variance
$ \widetilde{\,\sigma^2\,} = \sigma^2 / \langle M \rangle$, for which a Poisson distribution just gives unity. 

\begin{figure}[tbph!]
\center
\includegraphics[scale=0.35]{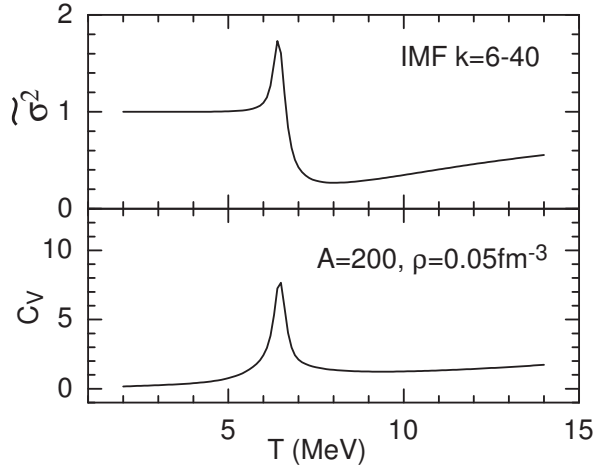}     % {m21-00.eps}
\caption{Specific heat $C_V$ and scaled variance $\widetilde{\sigma^2}$ as a function of temperature in one component canonical model. The temperature at maximum specific heat is defined as the phase transition temperature $T_b$ in the canonical model. For a system of size $A_{sys}=200$ and the density $\rho=0.05fm^{-3}$, the phase transition temperature is at $T_b=6.5$ MeV. The peak for maximum IMF correlations is at $T=6.4$ MeV. IMFs are defined as clusters of size $6 \leq k \leq 40$.} 
 \label{fig:comp1:variance-phase-transition}
\end{figure}

\begin{figure}[tbph!]
\center
\includegraphics[scale=0.35]{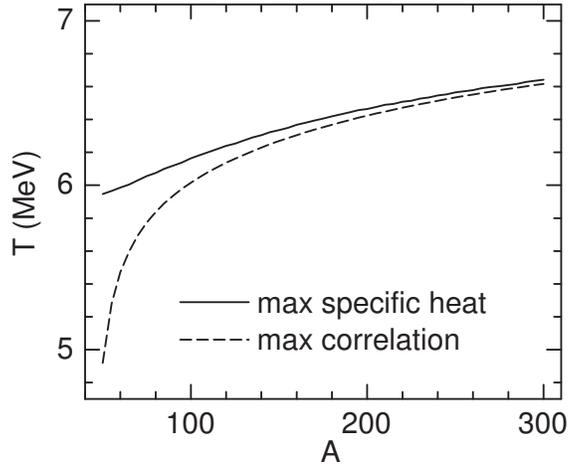}  %{m23-00.eps}
\caption{The temperatures for max specific heat and max correlation are plotted as a function of system size in the one component model. The density of the source is at $\rho / \rho_0 = 1/3$.}  
 \label{fig:comp1:two-different-T}
\end{figure} 

In the one component model, occurrence of maximum scaled variance (or simply maximum correlation) is simply related to the phase transition. As seen in Fig.\ref{fig:comp1:variance-phase-transition}, the phase transition temperature and the maximum variance almost occur at the same temperature. We can also see the close relation between temperatures for maximum correlation $T_c$ and for maximum specific heat $T_b$ in Fig.\ref{fig:comp1:two-different-T}. The difference between the two temperatures becomes smaller when the system size increases, and we could expect the two temperatures be the same for a infinite large system. Similar result for the peak and valley structure for variance has been reported for a lattice gas model \cite{Das:2000}.

\begin{figure}[tbph!]
\center
\includegraphics[scale=0.6]{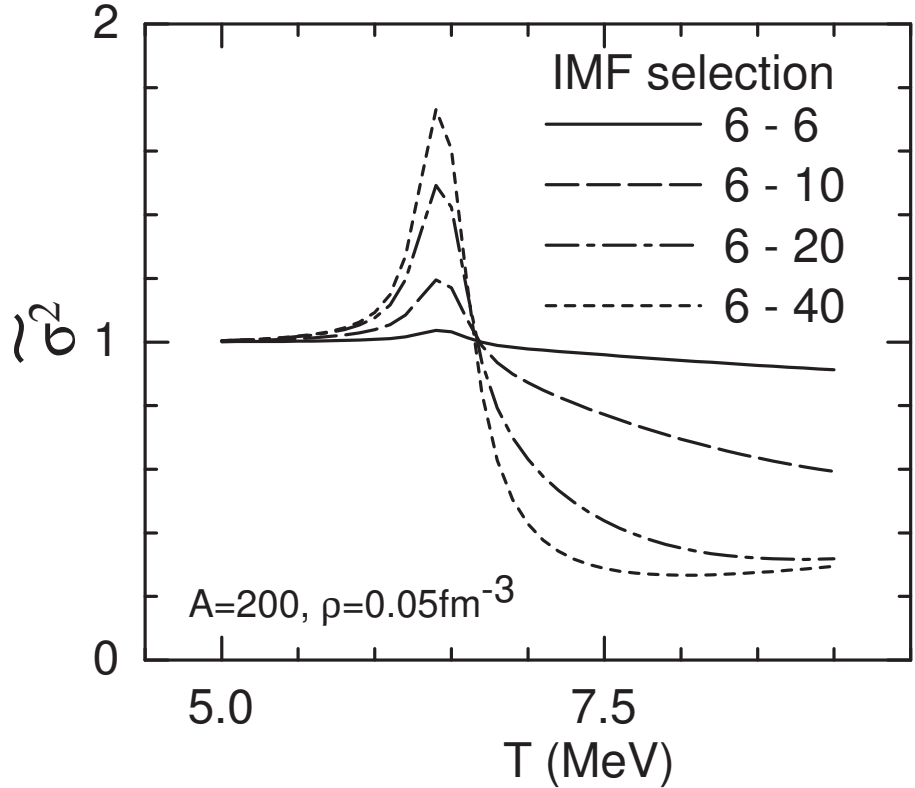}      %{m22-00.eps}
\caption{The scaled variance is plotted as a function of temperature in the phase transition region. Different lines correspond to different IMF selections. The change of the variance for a single cluster $k=6$ is quite small across critical temperature, while the change is significantly enhanced when including a large range of clusters. }  
 \label{fig:comp1:variance-build-up}
\end{figure}

As already observed in Sect.\ref{section:multiplicity:subsect:onecomp}, the multiplicity distributions shows systematic changes when the temperature of the system changes from below to above phase transition temperature. Plotted in Fig.\ref{fig:comp1:variance-build-up} is the scaled variance for multiplicity distributions of IMFs. The line labelled with "6-6" only selects clusters of size $k=6$, and the corresponding variance does not show a strong variation across the critical temperature region. Such a result is just as expected from our previous analysis on the multiplicity distributions of single clusters, where a simple Poisson distribution fits well the multiplicity distribution. The change of the variance for a single cluster is quite small, and poses difficulties when comparing with experiments.
However, as we select larger range of clusters, the variance shows significant structure of maximum and minimum in the region of phase transition temperature. The line with IMF selection of $6\leq k \leq 40$ shows a strong peak at close to the phase transition temperature, and a quick drop right above phase transition temperature. Such a behavior is an interesting feature, and is much easier to compare with experiments. The change of variance across the phase transition region has been observed before in a similar model \cite{Pratt:2000}.

The buildup of the variance is due to the correlations between clusters. If the clusters are uncorrelated, we have:
\begin{eqnarray}
\sigma^2 &=& \langle (\sum_i n_i)^2 \rangle - \langle (\sum_i n_i) \rangle^2  \,, \nonumber \\
&=& \sum_i \sigma_i^2 + \sum_{i,j} \left( \langle n_i n_j \rangle - \langle n_i \rangle \langle n_j \rangle  \right)
 \,, \nonumber \\ 
 &=&\sum_i \sigma_i^2 = \sum_i \langle n_i \rangle \widetilde{\,\sigma_i^2\,} \, ,     
  \label{eq:sect5:eq2}
\end{eqnarray}
where the cross terms are uncorrelated and cancel out. The scaled variance is 
\begin{eqnarray}
min(\widetilde{\,\sigma_i^2\,}) \leq \widetilde{\,\sigma^2\,} = \frac{\sigma^2}{\langle M \rangle} = \frac{\sum_i \langle n_i \rangle \widetilde{\,\sigma_i^2\,} }{\sum_i \langle n_i \rangle} \leq max(\widetilde{\,\sigma_i^2\,})
\,,       
  \label{eq:sect5:eq3}
\end{eqnarray}
Thus, the sum of uncorrelated distributions could not yield variance higher than the largest variance of the individual distributions, nor could it yield variance lower than the smallest variance of the individual distributions. Any significant enhancement or reduction of the variance as observed in Fig.\ref{fig:comp1:variance-build-up}, must come from correlations between clusters.

\begin{figure}[tbph!]
\center
\includegraphics[scale=0.6]{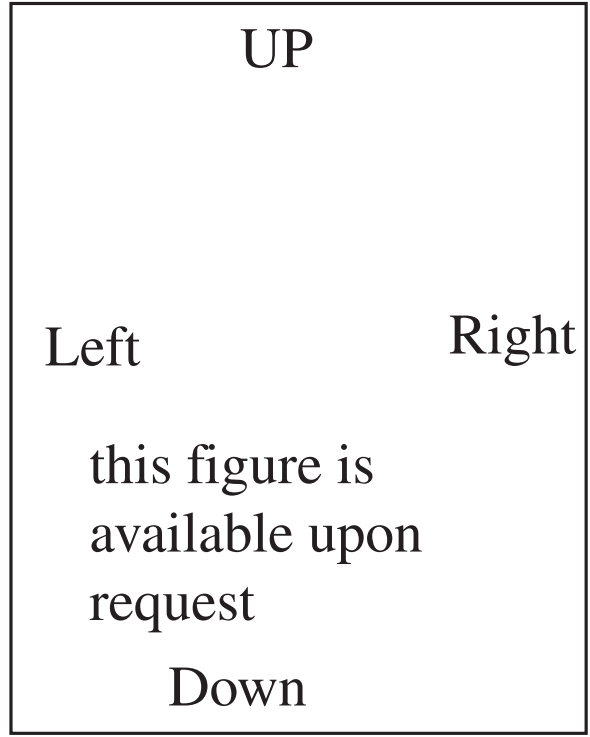}    %{m24-m0.eps}
\caption{The scaled variance as a function of temperature and system size for the one component canonical model. The shade of darkness represents the value of the scaled variance and the correspondence is shown by the insert scale. The contours regions are drawn in steps of $0.05$ in scaled variance, while contour lines are drawn in steps of $0.15$ starting from $\widetilde{\,\sigma^2\,}=0.275$.
The contour region corresponding to $\widetilde{\,\sigma^2\,}=0.6$ is darkened while that for $\widetilde{\,\sigma^2\,}=0.45$ is hatched. This result may be compared with Fig.3 in \cite{Beaulieu:1998a} in the transverse energy range $700 MeV \leq E_T \leq 1400 MeV$.}  
 \label{fig:comp1:variance-contour-Asys-T}
\end{figure}

As can be seen from Figs.\ref{fig:comp1:variance-build-up} and \ref{fig:comp1:variance-contour-Asys-T}, the variance peak is a continuous function of system size and temperature, so that it forms a ridge in the contour plot in Fig.\ref{fig:comp1:variance-contour-Asys-T}. The left hand side of the correlation ridge has variance always above one, and the right hand side are always below one, with a significant minimum at above but still close to the ridge. If experimentally we find the variance are below one and changes continuously with system size, the only allowable temperature in the one component model is above the phase transition temperature. If the variance value is larger than 0.2, which is the value for the minimum in this model, then there is practically only a very narrow range of allowable temperature just above the phase transition temperature.

We may compare the variance result with that from an experiment, Fig.3 in \cite{Beaulieu:1998a}. The region we are interested in are the more central collisions, with $700 MeV \leq E_T \leq 1400 MeV$, in $Xe+Au$ reactions at $110$ AMeV. The corresponding system size is estimated to be $150 \leq A_{sys} \leq 300$ assuming the linear relation between transverse energy and system size. The scaled variance changes from $\widetilde{\,\sigma^2\,}=0.6$ to $\widetilde{\,\sigma^2\,}=0.45$ in the data. The contour regions corresponding to these value of scaled variance are indicated in Fig.\ref{fig:comp1:variance-contour-Asys-T}. By projection of the hatched region to the temperature axis, we find only a narrow range of allowable temperature for the system. Such a tight constraint on the temperature is a unique feature of the one component model.

\begin{figure}[tbph!]
\center
\includegraphics[scale=0.6]{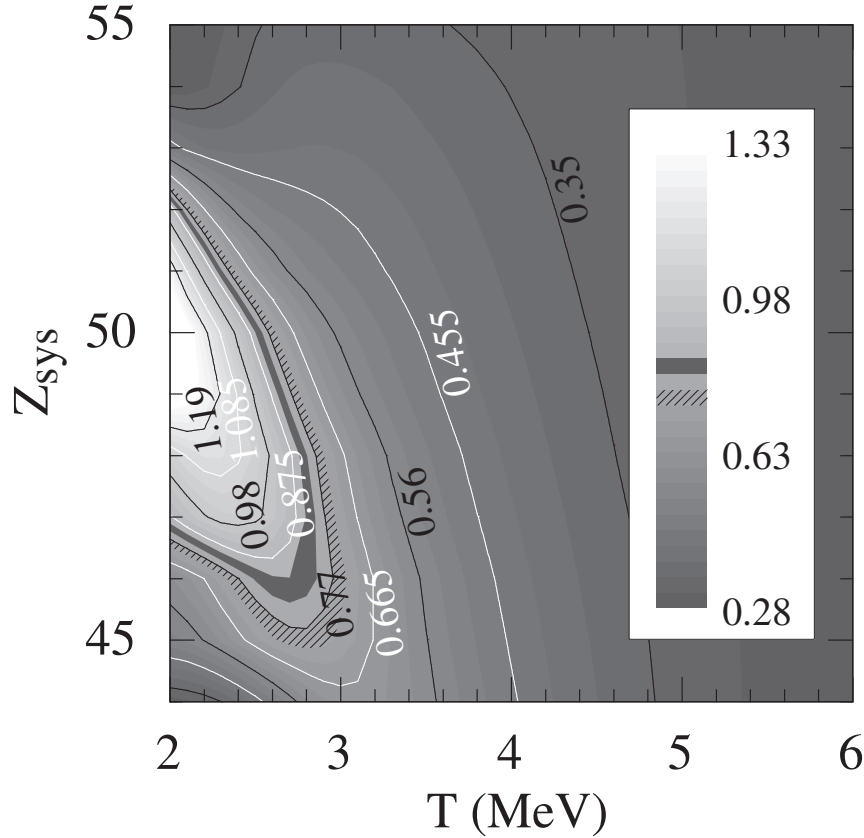}   %{m25-00.eps}
\caption{Contour plot of the scaled variance as a function of temperature and charge of the source $Z_{sys}$, where system size is fixed $A_{sys}=110$. The dark colored band indicates the region with $\widetilde{\,\sigma^2\,}=0.82$, corresponding to the variance measured by Moretto et al, in Ar+Au at 110 MeV/nuleon. The hashed band indicates the region with $\widetilde{\,\sigma^2\,}=0.75$, which is the decay corrected value for the variance. Each contour band corresponds to a change in variance by $0.035$, while contour lines are also drawn in steps of $0.105$ starting at $\widetilde{\,\sigma^2\,}=0.35$. The temperature obtained here is around $3$ MeV, if the isospin asymmetry of the source is around $\delta=0.15$. This temperature is slightly lower than $3.5$ MeV assumed from the multiplicity distribution result in Fig.\ref{fig:comp2:mdist-IMF-systemsize-25decay1} in Sect.\ref{sect:systemsize}.}  
 \label{fig:comp2:variance-3d-T-isospin-1}
\end{figure}

\begin{figure}[tbph!]
\center
\includegraphics[scale=0.6]{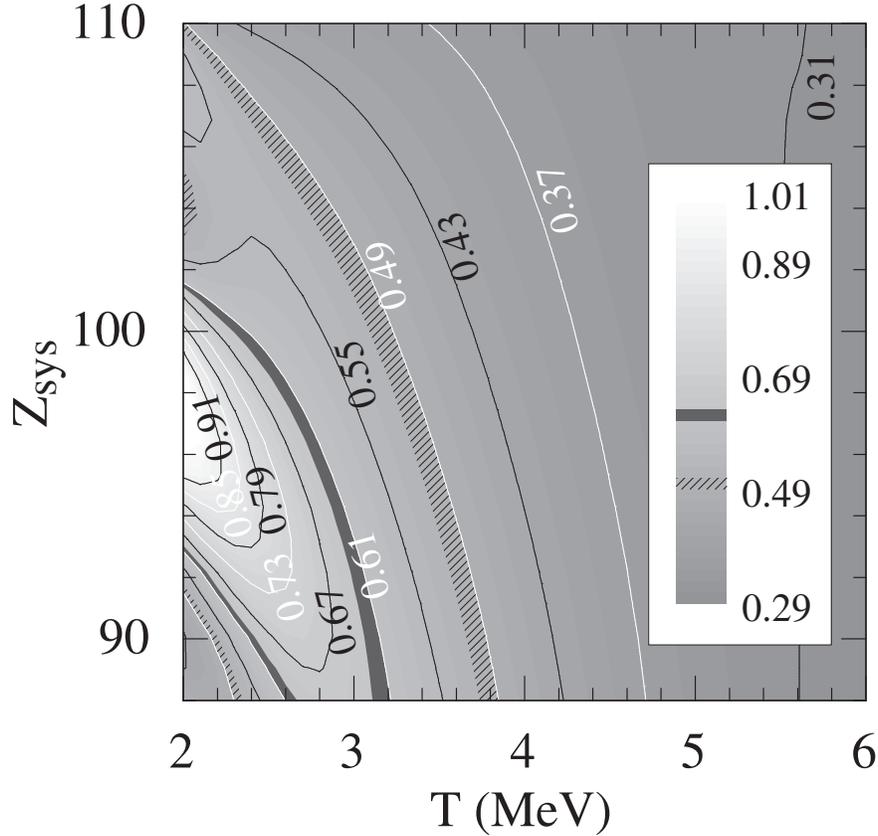}  %{m26-00.eps}
\caption{Contour plot of the scaled variance as a function of temperature $T$ and charge of the source $Z_{sys}$, where system size is fixed $A_{sys}=220$. The dark colored band indicates the region with $\widetilde{\,\sigma^2\,}=0.62$, corresponding to the variance measured by in Fig.4 in \cite{Moretto:1995}, in $Ar+Au$ at $110$ MeV/nuleon. The hashed band indicates the region with $\widetilde{\,\sigma^2\,}=0.5$, which is the decay corrected value for the variance. Each contour band corresponds to a change in variance by $0.02$, while contour lines are drawn in steps of $0.06$ starting from $\widetilde{\,\sigma^2\,}=0.31$. The temperature we find here is around $3.5$ MeV if the isospin asymmetry of the source is estimated as $\delta=0.15$. This temperature is in agreement with that from multiplicity distribution results in Sect.\ref{sect:systemsize}.}  
 \label{fig:comp2:variance-3d-T-isospin-2}
\end{figure}

For two component model, we have one additional degree of freedom in the isospin direction and situation became more complicated. The parameter space we are interested in is $\{ A,T,\delta \}$ here. We need to understand the variance dependence on isospin asymmetry $\delta$, or in other word, system charge $Z_{sys}$. Figs.\ref{fig:comp2:variance-3d-T-isospin-1} and \ref{fig:comp2:variance-3d-T-isospin-2} show the variance of IMFs for a system of $A_{sys}=110$ and 220 particles respectively. As can be seen, the isospin dependence of the variance is quite smaller in the higher temperature region. And there is a significant variance peak at low temperature of around $T=2 MeV$. It is generally believed that at very low temperature, a sequential emission scenario is more appropriate than the simultaneous fragmentation model \cite{Bondorf:1985b}. So we will try to look for regions where the temperature is not too low, generally $T \geq 3$ MeV.

We may compare the variance results with data from experiment. We will concentrate on the variance data as show in Fig.1 in \cite{Beaulieu:1998b}. Again, we will only consider more central collisions and assume simple linear relation between transverse energy and system size, specifically in the region of $550 \leq E_T \leq 1100 MeV$, or system size of about $110$ to $220$. The variance changes from $\widetilde{\,\sigma^2\,}=0.82$ to $\widetilde{\,\sigma^2\,}=0.62$, and the corresponding contour region are indicated with dark gray color in Figs.\ref{fig:comp2:variance-3d-T-isospin-1} and \ref{fig:comp2:variance-3d-T-isospin-2}. In the discussion in Sect. \ref{sect:systemsize}, we have assumed the a decay probability of $25\%$ to account for simple decay of IMFs and incomplete coverage in the detection of IMFs. As shown in Fig.\ref{fig:comp2:mdist-mean-variance-simpledecay}, the variance changes linearly with the decay probability, so we can make simple corrections for the decay process. The decay corrected variances are $\widetilde{\,\sigma^2\,}=0.75$ and $\widetilde{\,\sigma^2\,}=0.5$. The contour regions corresponding to the decay corrected variances are shown with hatched patterns in Figs.\ref{fig:comp2:variance-3d-T-isospin-1} and \ref{fig:comp2:variance-3d-T-isospin-2}. In the smaller system, a correction of decay only slightly increases the corresponding value for temperature, but the increase is more significant in the larger system. The extracted temperatures from Figs.\ref{fig:comp2:variance-3d-T-isospin-1} and \ref{fig:comp2:variance-3d-T-isospin-2} depend on the isospin asymmetry of the system, and we find around $T=3 MeV$ and $T=3.5 MeV$ for the two system if isospin asymmetry of the system is around $\delta=0.15$. In the smaller system, a significant change in isospin asymmetry of the system would yield considerably lower temperature, which we have tried to avoid. In the larger system, $A_{sys}=220$, the temperature we may get from pure decay corrected variance is in the range of $3 \leq T \leq 3.8$ MeV if the isospin of the system is in the range of $0.07 \leq \delta \leq 0.20$. The temperatures extracted for the larger system seem to agree well with the fitted temperature $T=3.5 MeV$ in Fig.\ref{fig:comp2:mdist-IMF-systemsize-25decay1}. But the smaller system does show some deviation. This discrepancy maybe due to the simple assumption of linear relation between transverse energy and system size, or the unrealistic assumption of constant temperature when transverse energy changes, or a combination of the two. As we have mentioned in Sect.\ref{sect:systemsize}, an increase system size, an increase in temperature and a reduction of decay probability have similar effects.    
In light of the isospin effect on variance in Fig.\ref{fig:comp2:mdist-IMF-systemsize-25decay1}, we might expect a slightly larger system size and a temperature of around 3 MeV to better fit both the multiplicity distribution result and the variance result at around $E_T = 550$ MeV.

\section{Summary}
\label{sect:summary}

In a canonical multifragmentation model, we used a recursive procedure to solve exactly the constrained partition problem. A recursive method for the multiplicity tagged partition function is derived. Multiplicity distributions of single or a range of clusters are calculated exactly. 

The Multiplicity distributions of single intermediate mass fragment (IMF) are found to be well approximated by a Poisson distribution. Details of the multiplicity distribution of single clusters show a critical behavior near the phase transition temperature in the one component canonical model. Following the suggestion of \cite{Beaulieu:1998a}, we defined a Poisson test parameter for the multiplicity distributions. This Poisson test parameter reveals details of the fragment production process near the phase transition. In a two component model, the multiplicity distribution of single clusters shows additional sensitivity to isospin asymmetry of the fragmentation source. The selection of a large range of clusters introduces strong correlations between clusters, thus the critical behavior shows up more strongly in the one component model. When comparing with experiments, the decay and feeding effects will significantly change the multiplicity distributions of single clusters or a range of clusters. However, the change of the mean multiplicity and the variance changes linearly with the decay probability. The production of single clusters is mostly uncorrelated, so that the multiplicity distribution of two clusters is well approximated by the simple product of each multiplicity distributions. 

The multiplicity distributions of single and a range of clusters have a smooth dependence on the system size. The crossings of different multiplicities have a strong correlation with system size if we assume the temperature is fixed. The multiplicity distribution of a range of clusters seems to show repeated patterns as a function of system size, which is very suggestive of repetition of similar fragmentation patterns as the system size increases. We can fit the data from \cite{Beaulieu:1998a,Moretto:1995} with simple assumption about the system size in a two component canonical model. An increase of system size, increase of temperature, and decrease of decay probability have similar effect on the multiplicity distribution. By requiring the system size fixed, we can find a single temperature that fits the data well. The Poisson test parameter for the distributions shows smooth dependence on system size. The Poisson test parameter for single clusters is always quite close to one, while the selection of a large range of clusters will generate more appreciable differences in the Poisson test parameter. 

The IMF multiplicity gated charge yield shows a change shape in the one component model, while such shape change did not show up in the two component model. The IMF multiplicity gated charge yield also shows correlations between IMF and other heavy clusters. Oscillating structure appeared in the zero IMF multiplicity gated yields, which is related to the rare process of binary fission type of fragmentation process, where the two fragments are strongly correlated. 

The variance is one of the most simple cumulants of the multiplicity distributions, and is also a simple measure of the correlations between IMFs. In a one component model, the critical behavior of the variance appears in the phase transition region. The selection of a large range of IMFs is shown to significantly enhance the change of variance across the phase transition region. The two component model is used to explain the correlation result from experiment \cite{Beaulieu:1998b}. The fitted parameters of temperature and decay probability are consistent with that from fitting the multiplicity distributions. 

Overall, we believe the multiplicity distributions and related signals provide new insights into the multifragmentation process in heavy-ion reactions. 
%%%%%%%%%%%%%%%%%%%%%%%%%%   
%template 
%
%\begin{eqnarray}
%\,,  \nonumber \\        
%  \label{eq:sect5:eq?}
%\end{eqnarray}
%
%\begin{figure}[tbph!]
%\center
%\includegraphics[scale=0.3]{box-9.eps}
%\caption{}  
% \label{}
%\end{figure}  
%%%%%%%%%%%%%%%%%%%%%%%%%%

\section{Acknowledgment}
This work is supported in part by the Natural Sciences and Engineering Research Council of Canada. 

%\newpage
\appendix
\section{multiplicity tagging}
\label{appendix:tag}

Let us denote each species that can emerge from the breakup as a,b,c,d,... {\em etc.} Let us divide these into two groups: $\alpha$ and $\beta$. A given species belongs to either $\alpha$ or $\beta$ but not to both. It is obvious that the partition function of the whole system which has A nucleons can be written as
\begin{eqnarray}
Q_A=\sum_{A_1,A_2} Q_{A_1}(\alpha) \, Q_{A_2}(\beta) \, \delta(A_1+A_2-A) \, .
 \label{eq:appendix:tagging:eq1}
\end{eqnarray}
Here $Q_{A_1}(\alpha)$ means the partition function of $A_1$ nucleons which can be partitioned into clusters in the $\alpha$ group. For example, $\alpha$ could comprise all the intermediate mass fragments. 

We want to do a multiplicity tagging on the partition within the $\alpha$ group. For this, we define
\begin{eqnarray}
 Q_{A_1}(\alpha, M) = \sum_{i \in \alpha} \, \prod_i \frac{(\omega_i)^{n_i}}{n_i !}  \, \delta(\sum i n_i - A_1 ) \, \delta(\sum n_i - M ) \, ,
 \label{eq:appendix:tagging:eq2}
\end{eqnarray} 
then clearly
\begin{eqnarray}
Q_{A}(\alpha)=\sum_M Q_{A} (\alpha, M) \, .
 \label{eq:appendix:tagging:eq3}
\end{eqnarray}
Starting from Eq.(\ref{eq:appendix:tagging:eq2}) and taking advantage of the identity
\begin{eqnarray}
\frac{1}{M}\sum_{k \in \alpha} n_k = 1 \, ,
 \label{eq:appendix:tagging:eq4}
\end{eqnarray}
we can build up a recursive formula for $Q_{A}(\alpha,M)$.
\begin{eqnarray}
 Q_{A}(\alpha, M) &=& \sum_{i \in \alpha} \left( \frac{1}{M}\sum_{k \in \alpha} n_k\right)\, \prod_i \frac{(\omega_i)^{n_i}}{n_i !}  \, \delta(\sum i n_i - A ) \, \delta(\sum n_i - M )   \nonumber \\
&=& \frac{1}{M} \sum_{i,\,k \in \alpha} n_k \frac{(\omega_k)^{n_k}}{n_k !}  \, \prod_{i \neq k} \frac{(\omega_i)^{n_i}}{n_i !} \, \delta(\sum i n_i + k n_k - A ) \, \delta(\sum n_i + n_k - M )   \nonumber \\
&=& \frac{1}{M} \sum_{i,\,k \in \alpha} \omega_k \frac{(\omega_k)^{n_k-1}}{(n_k-1) !} \, \prod_{i \neq k} \frac{(\omega_i)^{n_i}}{n_i !} \, \nonumber \\
&&\times \delta(\sum i n_i + k (n_k-1) - (A-k) ) \, \delta(\sum n_i + (n_k-1) - (M-1) )   \nonumber \\                                    
&=& \frac{1}{M} \sum_{k \in \alpha} \omega_k Q_{A-k}(\alpha, M-1) \, .
\label{eq:appendix:tagging:eq5}
\end{eqnarray} 
Going back to the full system $(\alpha+\beta)$, the relevant probability is given by
\begin{eqnarray}
P(\alpha,M) = \frac{1}{Q_A} \sum_{A_1,A_2} Q_{A_1}(\alpha,M) Q_{A_2}(\beta)  \delta(A_1+A_2-A) \,.
  \label{eq:appendix:tagging:eq6}
\end{eqnarray}

\section{Decay and feeding}
\label{appendix:decay}

We may start from the simple case of decay only problem. To make it even simpler, we will assume the clusters in the group $\alpha$ have equal probability $\epsilon$ to decay into clusters outside the group $\alpha$. The decay of clusters within the group $\alpha$ is not considered since they do not change the multiplicity distribution.

If before decay, the multiplicity distribution is $P(\alpha, N)$, then after independent decay of each of these cluster in the group $\alpha$, the multiplicity $N$ event will become multiplicity $M$ event ($M=0,1,2,...,N$), with a weight of
\begin{eqnarray}
P(N \rightarrow M) = C_M^N \epsilon^{N-M} (1-\epsilon)^M \,,
  \label{eq:appendix:decay:eq1}
\end{eqnarray} 
Then the probability of multiplicity M event after decay is
\begin{eqnarray}
P_{d}(\alpha,N \rightarrow M)= \sum_{N} P(\alpha,N) P(N \rightarrow M) \,,
  \label{eq:appendix:decay:eq2}
\end{eqnarray}

However, when the decay probability of the clusters in group $\alpha$ is not uniform,
the above equation will not work. For this more general case, we have to tag the multiplicity of each of the clusters in the group $\alpha$, which will be a formidable task for even a moderate sized system. 

%Consider that we can tagged each of the cluster $i$ within the group $\alpha$. 
%The multiplicity tagged formation probability is $P(i,n_i)$, and such a 
%multiplicity $n_i$ event became multiplicity $N_i$ event with a weight

Because of decay, if the number of composite of type $i$ in group $\alpha$ was $n_i$, this can turn into $N_i$ composites in group $\alpha$ with probability 
\begin{eqnarray}
P(n_i \rightarrow N_i) = C_{N_i}^{n_i} \epsilon_i^{n_i-N_i} (1-\epsilon_i)^{N_i}\,.
  \label{eq:appendix:decay:eq3}
\end{eqnarray}

The total after decay tagged partition function is:
\begin{eqnarray}
Q_A^D(\alpha,M) &=& \sum \prod_{i \in \alpha} \frac{(\omega_i)^{n_i}}{n_i !} P(n_i \rightarrow N_i) \prod_{j \notin \alpha} \frac{(\omega_j)^{n_j}}{n_j !} \, 
 \nonumber \\ 
&& \times \delta(\sum_i i n_i + \sum_j j n_j -A)\, \delta(\sum_i N_i - M)  \,,
  \label{eq:appendix:decay:eq4}
\end{eqnarray}  

The total weight factor for cluster $i$ can be simplified:
\begin{eqnarray}
 \frac{(\omega_i)^{n_i}}{n_i !} P(n_i \rightarrow N_i) &=& C_{N_i}^{n_i} \epsilon_i^{n_i-N_i} (1-\epsilon_i)^{N_i} \frac{(\omega_i)^{n_i}}{n_i !} \,, 
 \nonumber \\
&=& \frac{(\epsilon_i \omega_i)^{n_i-N_i}}{(n_i-N_i)!} \frac{[(1-\epsilon_i)(\omega_i)]^{N_i}}{N_i !}  \, .
  \label{eq:appendix:decay:eq5}
\end{eqnarray}  
Here we may view the first term as a partition weight for a new type of cluster $i'$, where the formation probability is $\tilde{\omega}_i'=(\epsilon_i \omega_i)$, $N_i'=n_i-N_i$; and the second term in the partition could be viewed as a weight for the after decay cluster $i$, with formation probability $\tilde{\omega}_i=[(1-\epsilon_i)(\omega_i)]$. With this, 
\begin{eqnarray}
\frac{(\omega_i)^{n_i}}{n_i !} P(n_i \rightarrow N_i) \mapsto 
\frac{(\tilde{\omega}_i')^{N_i'}}{N_i' !} \frac{(\tilde{\omega}_i)^{N_i}}{N_i ! }
\,,   
  \label{eq:appendix:decay:eq6}
\end{eqnarray} 
We will denote the after decay clusters $i \in \alpha$ as the stay group $\alpha^S$, and the after decay clusters $i'$ as a new group $\alpha^D$.

With the help of Eq.(\ref{eq:appendix:decay:eq6}), the total tagged partition Eq.(\ref{eq:appendix:decay:eq4}) is simplified into
\begin{eqnarray}
Q_A^D(\alpha,M) 
&=& \sum \prod_{i \in \alpha^S} \frac{(\tilde{\omega}_i)^{N_i}}{N_i !}  \prod_{i' \in \alpha^D} \frac{(\tilde{\omega}_i')^{N_i'}}{N_i' !} \prod_{j \notin \alpha^S, j\notin \alpha^D} \frac{(\tilde{\omega}_j)^{N_j}}{N_j !} \,, 
  \nonumber \\    
&& \times \delta(\sum_{i} N_i - M)\, \delta(\sum_{i} i N_i +\sum_{i'} i' N_i' +\sum_{j} j N_j- A)  
  \label{eq:appendix:decay:eq7}
\end{eqnarray} 
Where the new formation probability is defined as
\begin{eqnarray}
\tilde{\omega_i} = \left\{
    \begin{array}{cl}
     (1-\epsilon_i) \omega_i,      & \mbox{$i \in \alpha^S$}; \\
     \epsilon_i \omega_i,      & \mbox{$i \in \alpha^D$}; \\
     \omega_i,      & \mbox{$i \notin \alpha^S, i \notin \alpha^D$}.
    \end{array}
  \right.     
  \label{eq:appendix:decay:eq8}
\end{eqnarray} 
Since we are only interested in the group $\alpha^S$, where tagging is needed, the rest of the clusters could be combined into a new group called $\beta'=\alpha^D+other$. 
\begin{eqnarray}
Q_A^D(\alpha,M) 
 &=& \sum 
    \prod_{i \in \alpha^S} \frac{(\tilde{\omega}_i)^{N_i}}{N_i !}
    \delta(\sum_{i \in \alpha^S} N_i - M) \delta(\sum_{i \in \alpha^S} i N_i -A_1)
 \,,  \nonumber \\
 && \times \prod_{j \in \beta'} \frac{(\tilde{\omega}_j)^{N_j}}{N_j !}
 \delta(\sum_{j \in \beta'} j N_j- A_2)  
 \, \delta(A_1 +A_2- A) 
 \,,  \nonumber \\
 &=& \sum_{A_1,A_2} Q_{A_1}(\alpha^S,M) Q_{A_2} (\beta') \, \delta(A_1 +A_2- A) \, .
 \label{eq:appendix:decay:eq9}
\end{eqnarray}
And the multiplicity distribution is now
\begin{eqnarray}
P^D(\alpha,M) = \frac{1}{Q_A} \sum_{A_1,A_2} Q_{A_1}(\alpha^S,M) Q_{A_2} (\beta') 
  \, \delta(A_1 +A_2- A)        
  \label{eq:appendix:decay:eq10}
\end{eqnarray} 
 
Notice Eq.(\ref{eq:appendix:decay:eq10}) resembles Eq.(\ref{eq:appendix:tagging:eq6}), the differences are in the grouping of clusters, and in the formation probability ($\tilde{\omega}$ instead of $\omega$).

The decay of clusters in group $\beta$ could also feed into the group $\alpha$, and therefore changes the multiplicity distribution of after-decay multiplicity distribution of the clusters in group $\alpha$. Since each of these clusters decays independently with a different decay probability, we have to tag each of these before-decay clusters in group $\beta$ as well as $\alpha$. The multiplicity for the cluster $(j,n_j)$ will be multiplied by the decay weight factor
\begin{eqnarray}
P(n_j \rightarrow N_j) = C_{N_j}^{n_j} \epsilon_j^{N_j} (1-\epsilon_j)^{n_j-N_j}\,.
  \label{eq:appendix:decay:eq11}
\end{eqnarray}
Notice Eq.(\ref{eq:appendix:decay:eq11}) differs from Eq.(\ref{eq:appendix:decay:eq4}) by only the exchange of $\epsilon \leftrightarrow (1-\epsilon)$.

The total after-decay tagged partition function is
\begin{eqnarray}
Q_A^{D,F}(\alpha,M) 
 &=& \sum \prod_{i \in \alpha} \frac{(\omega_i)^{n_i}}{n_i !} P(n_i \rightarrow N_i) \prod_{j \in \beta} \frac{(\omega_j)^{n_j}}{n_j !}P(n_j \rightarrow N_j) 
 \prod_{k \notin \alpha, k \notin \beta} \frac{(\omega_k)^{n_k}}{n_k !}\, 
 \nonumber \\ 
&& \times \delta(\sum_i i N_i + \sum_j j N_j +\sum_k k N_k-A)\, \delta(\sum_i N_i +\sum_j N_j - M)  \,,
  \label{eq:appendix:decay:eq12}
\end{eqnarray} 
As can be readily seen, the factorization of Eq.(\ref{eq:appendix:decay:eq6}) works for both $i \in \alpha$ and $j \in \beta$.
We only need a redefinition of the formation probability $\omega$
\begin{eqnarray}
\tilde{\omega_i} = \left\{
    \begin{array}{cl}
     (1-\epsilon_i) \omega_i,      & \mbox{$i \in \alpha^S$}; \\
     \epsilon_i \omega_i,      & \mbox{$i \in \alpha^D$}; \\
     (1-\epsilon_i) \omega_i,      & \mbox{$i \in \beta^S$};  \\
     \epsilon_i \omega_i,      & \mbox{$i \in \beta^D$}; \\
     \omega_i,      & \mbox{$i \notin \alpha$, $i \notin \beta$}. 
    \end{array}
  \right.     
  \label{eq:appendix:decay:eq13}
\end{eqnarray}
   
And the tagged after-decay partition function is 
\begin{eqnarray}
Q_A^{D,F}(\alpha,M) &=& \sum 
 \prod_{i \in \alpha^S} \frac{(\tilde{\omega}_i)^{N_i}}{N_i !}
 \prod_{k \in \beta^D} \frac{(\tilde{\omega}_k)^{N_k}}{N_k !}
  \delta(\sum_{i \in \alpha^S} N_i +\sum_{k \in \beta^D} N_k- M)\,, 
  \nonumber \\                                                                  
&& \times  \,
 \prod_{j \in \alpha^D} \frac{(\tilde{\omega}_j)^{N_j}}{N_j !} 
 \prod_{l \in \beta^S} \frac{(\tilde{\omega}_l)^{N_l}}{N_l !} 
 \prod_{m \notin \alpha, m\notin \beta} \frac{(\tilde{\omega}_m)^{N_m}}{N_m !} \,, 
  \nonumber \\    
&& \times  \,
 \delta(\sum_{i } i N_i +\sum_{j} j N_j +\sum_{k} k N_k+\sum_{l} l N_l+\sum_{m} m N_m- A)  
  \label{eq:appendix:decay:eq14}
\end{eqnarray}      
Define the new set of interest $\alpha'=\alpha^S+\beta^D$, and irrelevant $\beta'=\alpha^D+\beta^S+other$. Then we have
\begin{eqnarray}
Q_A^{D,F}(\alpha,M) &=& \sum 
 Q_{A_1}(\alpha',M) Q_{A_2}(\beta') \, \delta(A_1+A_2-A) \, .
 \label{eq:appendix:decay:eq15}
\end{eqnarray}

%\newpage

\end{document}